\documentclass{emulateapj}
\usepackage{times}
\usepackage{graphicx}

\shorttitle{\emph{Fermi}-LAT Observations of the Crab Pulsar and Nebula}
\shortauthors{A.A. Abdo et al.}

\begin{document}


\title{\emph{Fermi} Large Area Telescope Observations\\
	of the Crab Pulsar and Nebula}


\author{
A.~A.~Abdo\altaffilmark{2,3}, 
M.~Ackermann\altaffilmark{4}, 
M.~Ajello\altaffilmark{4}, 
W.~B.~Atwood\altaffilmark{5}, 
M.~Axelsson\altaffilmark{6,7}, 
L.~Baldini\altaffilmark{8}, 
J.~Ballet\altaffilmark{9}, 
G.~Barbiellini\altaffilmark{10,11}, 
M.~G.~Baring\altaffilmark{12}, 
D.~Bastieri\altaffilmark{13,14}
% ...
}
% ...
\altaffiltext{2}{Space Science Division, Naval Research Laboratory, Washington, DC 20375, USA}
\altaffiltext{3}{National Research Council Research Associate, National Academy of Sciences, Washington, DC 20001, USA}
\altaffiltext{4}{W. W. Hansen Experimental Physics Laboratory, Kavli Institute for Particle Astrophysics and Cosmology, Department of Physics and SLAC National Accelerator Laboratory, Stanford University, Stanford, CA 94305, USA}
\altaffiltext{5}{Santa Cruz Institute for Particle Physics, Department of Physics and Department of Astronomy and Astrophysics, University of California at Santa Cruz, Santa Cruz, CA 95064, USA}
\altaffiltext{6}{Department of Astronomy, Stockholm University, SE-106 91 Stockholm, Sweden}
\altaffiltext{7}{The Oskar Klein Centre for Cosmoparticle Physics, AlbaNova, SE-106 91 Stockholm, Sweden}
\altaffiltext{8}{Istituto Nazionale di Fisica Nucleare, Sezione di Pisa, I-56127 Pisa, Italy}
\altaffiltext{9}{Laboratoire AIM, CEA-IRFU/CNRS/Universit\'e Paris Diderot, Service d'Astrophysique, CEA Saclay, 91191 Gif sur Yvette, France}
\altaffiltext{10}{Istituto Nazionale di Fisica Nucleare, Sezione di Trieste, I-34127 Trieste, Italy}
\altaffiltext{11}{Dipartimento di Fisica, Universit\`a di Trieste, I-34127 Trieste, Italy}
\altaffiltext{12}{Rice University, Department of Physics and Astronomy, MS-108, P. O. Box 1892, Houston, TX 77251, USA}
\altaffiltext{13}{Istituto Nazionale di Fisica Nucleare, Sezione di Padova, I-35131 Padova, Italy}
\altaffiltext{14}{Dipartimento di Fisica ``G. Galilei", Universit\`a di Padova, I-35131 Padova, Italy}
% ...

\begin{abstract}
We report on $\gamma$-ray observations of the Crab Pulsar and Nebula using 8 months of survey data with the \emph{Fermi} Large Area Telescope (LAT). The high quality light curve obtained using the ephemeris provided by the Nan\c{c}ay and Jodrell Bank radio telescopes shows two main peaks stable in phase with energy. The first $\gamma$-ray peak leads the radio main pulse by (281 $\pm$ 12 $\pm$ 21) $\mu$s, giving new constraints on the production site of non-thermal emission in pulsar magnetospheres. The first uncertainty is due to $\gamma$-ray statistics, and the second arises from the rotation parameters. The improved sensitivity and the unprecedented statistics afforded by the LAT enable precise measurement of the Crab Pulsar spectral parameters: cut-off energy at $E_{c}$~=~($5.8~\pm~0.5~\pm~1.2$)~GeV, spectral index of $\Gamma$~=~($1.97~\pm~0.02~\pm~0.06$) and integral photon flux above 100 MeV of $(2.09~\pm~0.03~\pm~0.18$)~$\times$~10$^{-6}$ cm$^{-2}$~s$^{-1}$. The first errors represent the statistical error on the fit parameters, while the second ones are the systematic uncertainties. Pulsed $\gamma$-ray photons are observed up to $\sim$~20 GeV which precludes emission near the stellar surface, below altitudes of around 4 to 5 stellar radii in phase intervals encompassing the two main peaks. A detailed phase-resolved spectral analysis is also performed: the hardest emission from the Crab Pulsar comes from the bridge region between the two $\gamma$-ray peaks while the softest comes from the falling edge of the second peak. The spectrum of the nebula in the energy range 100 MeV -- 300 GeV is well described by the sum of two power-laws of indices $\Gamma_{sync}$~=~($3.99~\pm~0.12~\pm~0.08$) and $\Gamma_{IC}$~=~($1.64~\pm~0.05~\pm~0.07$), corresponding to the falling edge of the synchrotron and the rising edge of the inverse Compton components, respectively. This latter, which links up naturally with the spectral data points of Cherenkov experiments, is well reproduced via inverse Compton scattering from standard Magnetohydrodynamics (MHD) nebula models, and does not require any additional radiation mechanism.
\end{abstract}

\keywords{gamma rays: observations -- pulsars: individual (Crab, PSR J0534+2200) -- ISM: supernova remnants -- ISM: individual: Crab Nebula -- Fermi}

\sloppy

\section{Introduction}
The Crab Nebula belongs to the class of filled-center supernova remnants (SNR) \citep{Green 2006}, i.e. without any detected shell component, and is well studied in almost all wavelength bands of the electromagnetic spectrum from the radio ($10^{-5}$~eV) to very high energy $\gamma$-rays (nearly $10^{14}$~eV). It is held to be the archetypical pulsar wind nebula, even though its physical and spectral properties are unique. It is associated with the supernova explosion reported by Chinese astronomers in 1054 AD. Several models (\cite{Kennel and Coroniti 1984}, \cite{de Jager and Harding 1992}, \cite{de Jager et al. 1996} and references therein) describe the photon production processes taking place in this nebula. Synchrotron radiation from high energy electrons in the nebular magnetic field is responsible for the observed spectrum from radio to MeV, while inverse Compton (IC) scattering of the primary accelerated electrons off the synchrotron photons, far infrared and Cosmic Microwave Background (CMB) produces high energy $\gamma$-rays. While these two mechanisms seem to provide a reasonable description of the overall non-thermal radiation of the Crab Nebula, one cannot exclude possible deviations from this simplified picture, and \cite{Atoyan and Aharonian 1996} proposed that significant production of high energy $\gamma$-rays by bremsstrahlung radiation of relativistic electrons could take place in the Crab filaments. This paper, reporting the results of a precise spectral analysis of the Crab Nebula between 100~MeV and 300~GeV, adds new elements to this discussion.

At the center of the nebula lies the Crab Pulsar, PSR~J0534+2200, one of the most energetic known pulsars (spin down power of $\dot{E}$ = 4.6 $\times$ 10$^{38}$ erg~s$^{-1}$), located at a distance of (2.0 $\pm$ 0.2) kpc. Estimation of its characteristic age using its rotation period ($P~=~33$~ms) and derivative ($\dot{P}~=~4.2 \times 10^{-13}$~s/s) yields an age of 1240 years, close to the observational value. 

The Energetic Gamma-Ray Experiment Telescope (EGRET), on orbit from 1991 to 2001 on board of the Compton Gamma Ray Observatory (CGRO), reported the high energy detection of the Crab Nebula and Pulsar \citep{Nolan et al. 1993, de Jager et al. 1996, Kuiper et al. 2001}. A more detailed study of the $\gamma$-ray emission was then provided by \cite{Fierro et al. 1998}, presenting a complete phase-resolved spectral analysis of the EGRET data, and more recently by \cite{Pellizzoni et al. 2009}, describing the first AGILE timing results on $\gamma$-ray pulsars, including the Crab. Pulsations of the Crab Pulsar were reported above 25 GeV by the Major Atmospheric Gamma-ray Imaging Cherenkov Telescope (MAGIC) collaboration \citep{Aliu et al. 2008}, with a light curve consistent with the one measured by EGRET. 

Observations of the Crab Pulsar in high energy $\gamma$-rays can provide strong constraints on the location of the $\gamma$-ray emitting regions: above the polar caps \citep{Daugherty and Harding 1996}, in the intermediate models like the slot gap \citep{Muslimov and Harding 2004}, or far from the neutron star in the outer gaps \citep{Romani 1996}. In particular, the spectral analysis and the phase-resolved behaviour examined in this paper may be used to discriminate between these models.

Successfully launched on June 11, 2008, the Large Area Telescope (LAT), aboard the \emph{Fermi} Gamma-ray Space Telescope, formerly GLAST, offers the unique opportunity to study the high energy behaviour of the Crab Pulsar and Nebula in great detail. In this paper, we report the results of the analysis of the Crab region using 8 months of survey observations with the \emph{Fermi}-LAT. In Sections~\ref{radio} and \ref{lat}, we describe the radio and $\gamma$-ray observations used, while Section~\ref{results} presents the results obtained from a detailed timing and spectral analysis of the LAT data. Finally, in Sections~\ref{discussion} and \ref{summary}, we discuss and summarize the main implications of these results for models of both the pulsar and the nebula. 

\section{Radio Timing observations}
\label{radio}
The Crab Pulsar, like many young pulsars, is affected by significant timing noise and glitches in rotation rate. Because of the long time interval considered in this article, the rotational phase behaviour with time has to be known with extreme precision. The Crab Pulsar is one of the more than two hundred pulsars of large spin-down luminosity $\dot{E}$ monitored by the LAT pulsar timing campaign \citep{Smith et al. 2008} coordinated among \emph{Fermi}, radio and X-ray telescopes.

The timing solution for the Crab Pulsar has been built using observations made with the Nan\c{c}ay radio telescope (France) \citep{Theureau et al. 2005} and the Jodrell Bank Observatory 42 foot MKIA telescope (England) \citep{Hobbs et al. 2004, Lyne et al. 1993}. A total of 698 times of arrival (TOAs) has been recorded between June 20, 2008 and April 8, 2009. The radio TOA dataset comprises 210 observations at 1.4 GHz and 488 observations at 600 MHz, in order to constrain the dispersion measure during the interval of the $\gamma$-ray observations. This quantity is known to be highly variable, because of moving filaments in the Crab Nebula which make the column density of electrons along the line-of-sight change in time.

The TEMPO2 timing package \citep{Hobbs et al. 2006} was used to build the timing solution from the 698 TOAs. The mean time of arrival uncertainties are 2.66 and 15.64 $\mu$s for the Nan\c{c}ay and the Jodrell Bank Observatory observations, respectively. We fit the TOAs to the pulsar rotation frequency and its first two derivatives, as well as to the dispersion measure (DM) and its first derivative (DM1) to take the variation of the electron column density into account. The fit further includes 10 harmonically-related sinusoids,
using the ``FITWAVES'' option in the TEMPO2 package, to flatten the timing noise. We obtain DM~=~(56.7037~$\pm$~0.0003) cm$^{-3}$~pc and DM1~=~(3.05~$\pm$~0.10)~$\times$~10$^{-2}$ cm$^{-3}$~pc~yr$^{-1}$ on November 11, 2008. The post-fit rms is 21.1 $\mu$s, allowing for analyses of the $\gamma$-ray pulse profile with unprecedented precision.
\section{LAT description and observations}
\label{lat}
The LAT is an electron-positron pair conversion telescope, sensitive to $\gamma$-rays with energies from below 20 MeV to more than 300 GeV. It consists of a high-resolution converter tracker (direction measurement of the incident $\gamma$-rays), a CsI(Tl) crystal calorimeter (energy measurement) and an anticoincidence detector to discriminate the background of charged particles \citep{Atwood et al. 2009}. In comparison to its predecessor EGRET, the LAT has a larger effective area ($\sim$ 8000 cm$^{2}$ on-axis), a broader field of view ($\sim$ 2.4 sr) and a superior angular resolution ($\sim$ 0.6$^{\circ}$ 68$\%$ containment at 1 GeV for events converting in the front section of the tracker).

The following analysis was performed on 248 days of data taken in survey mode (August 02, 2008 -- April 07, 2009). Events from the "Diffuse" class are selected, i.e. the highest quality photon data, having the most stringent background rejection. In addition, we exclude the events with zenith angles greater than 105$^{\circ}$ due to the Earth's bright $\gamma$-ray albedo. 

\section{Results}
\label{results}
\subsection{Light curves}
\label{phasos}
The selected $\gamma$-rays were phase-folded using the timing solution described in Section~\ref{radio}. Photons with an angle $\theta< \rm{Max}(6.68 - 1.76 \rm{Log_{10}(E_{MeV}}),1.3)^{\circ}$ of the radio pulsar position, R.A.~=~83.63322$^{\circ}$, Dec.~=~22.01446$^{\circ}$ (J2000), are selected. This choice takes into account the instrument performance and maximizes the signal to noise ratio. At high energies, the background is relatively faint compared to the Crab emission, so that a radius larger than the Point Spread Function (PSF) can be kept.

\begin{figure*}[h!]
\epsscale{1.10}
\plotone{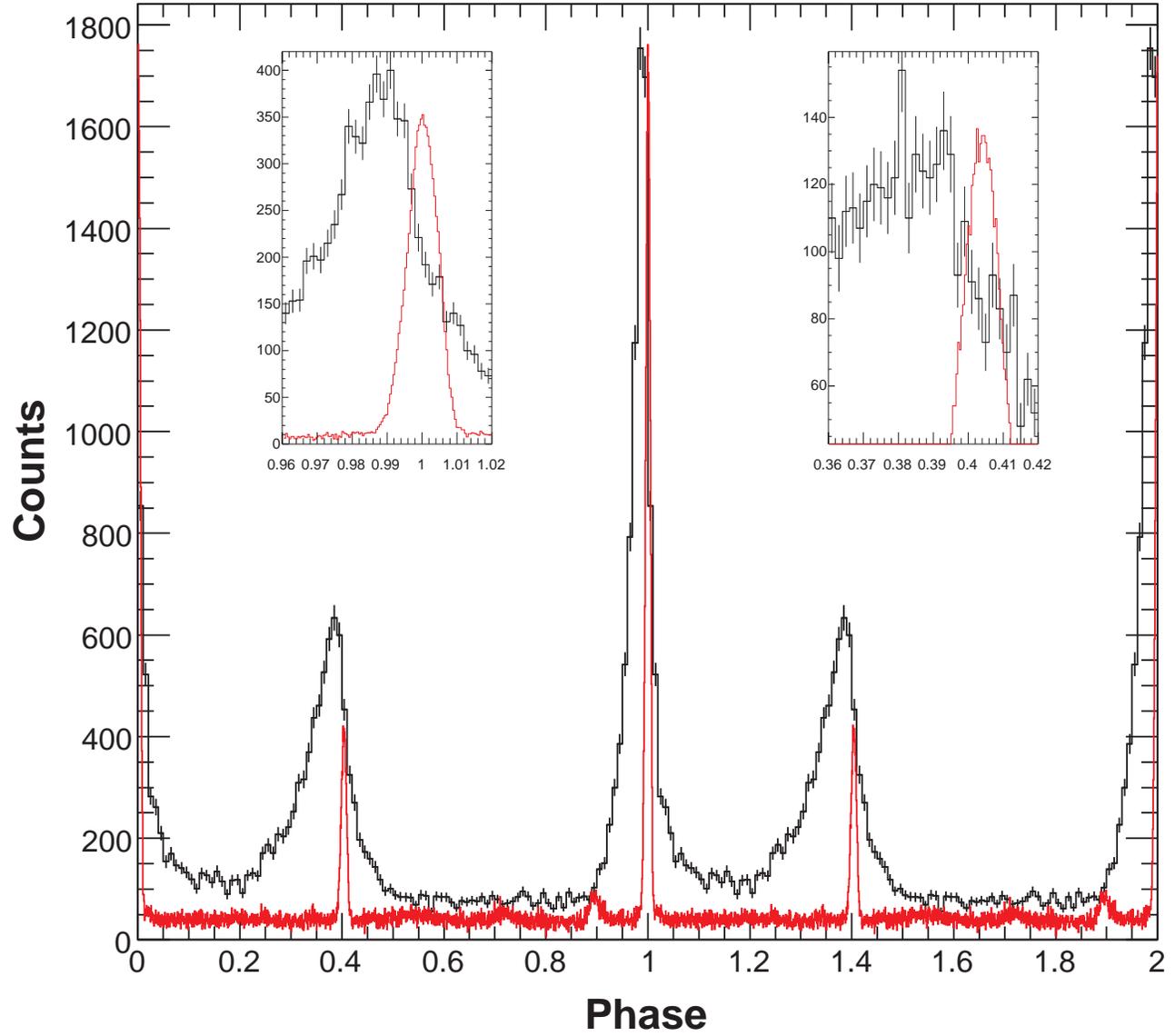}
\caption{\label{phaso_general}Light curve obtained with photons above 100 MeV within an energy-dependent circular region, as described in Section \ref{phasos}. The light curve profile is binned to 0.01 of pulsar phase. Insets show the pulse shapes near the peaks, binned to 0.002 in phase. The radio light curve (red line) is overlaid (arbitrary units). The main peak of the radio pulse seen at 1.4 GHz is at phase 0. Two cycles are shown.}
\end{figure*}

\begin{figure*}
\epsscale{1.1}
\plotone{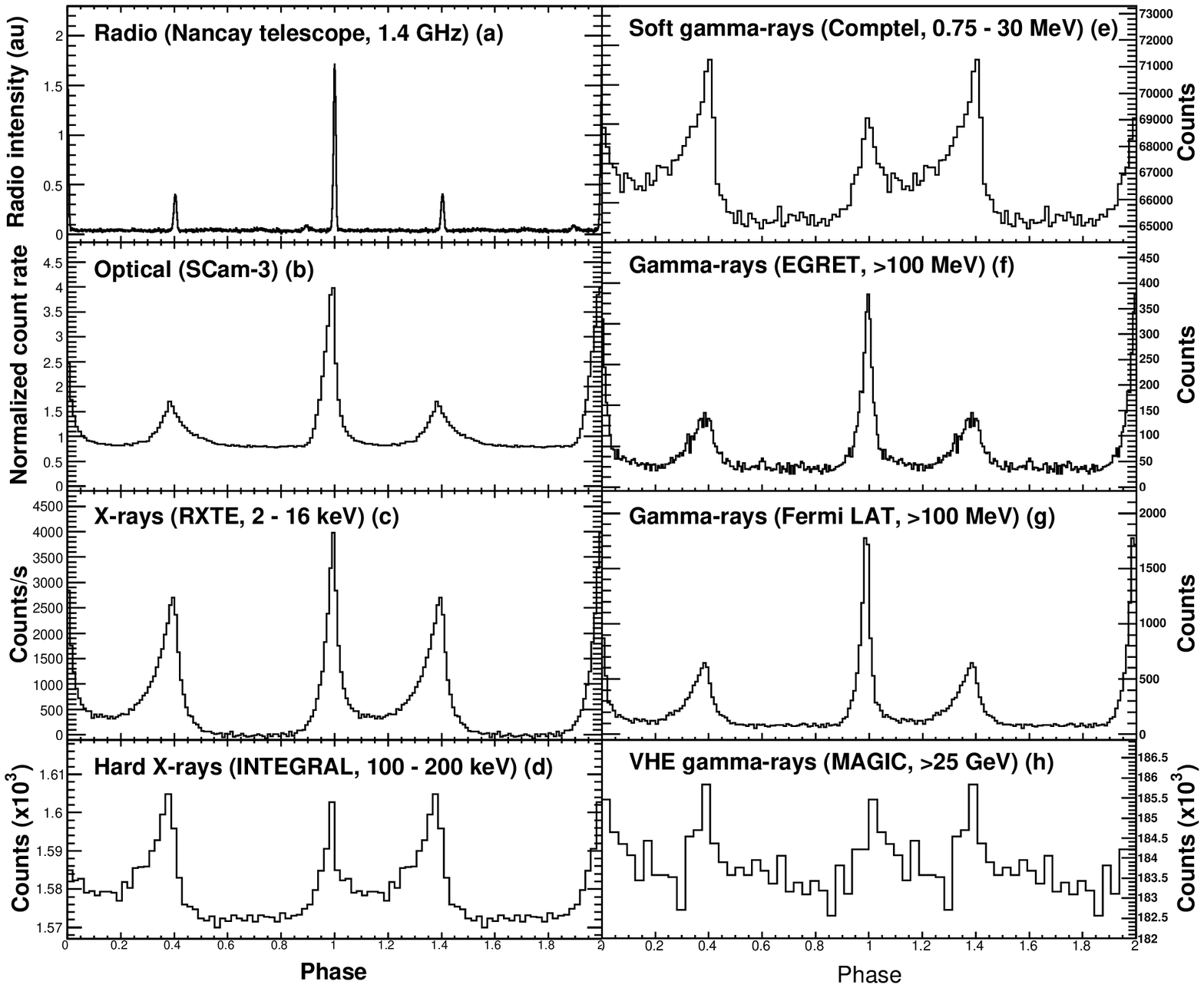}
\caption{\label{phaso_multilambda}Light curves at different wavelengths. Two cycles are shown. References: (a) from the Nan\c{c}ay radio telescope; (b) \cite{Oosterbroek et al. 2008}; (c) \cite{Rots et al. 2004}; (d) \cite{Mineo et al. 2006}; (e) \cite{Kuiper et al. 2001}; (f) EGRET, \cite{Kuiper et al. 2001}; (g) this paper; (h) \cite{Aliu et al. 2008}.}
\end{figure*}

Using this energy-dependent region, the $\gamma$-ray light curve above 100 MeV is presented in Figures~\ref{phaso_general} and \ref{phaso_multilambda} (g). We have 22601 $\gamma$-rays among which we estimate 14563 $\pm$ 240 pulsed photons after background subtraction. Phase histograms in radio (from the Nan\c{c}ay radio telescope), optical \citep{Oosterbroek et al. 2008}, X-rays \citep{Rots et al. 2004}, hard X-rays \citep{Mineo et al. 2006}, $\gamma$-rays (CGRO COMPTEL and EGRET, \cite{Kuiper et al. 2001}) and very high energy (VHE) $\gamma$-rays (MAGIC, \cite{Aliu et al. 2008}) are also plotted in Figure~\ref{phaso_multilambda}. We did not search for any correlation between giant pulses and $\gamma$-ray photons.

\begin{figure*}[h!]
\epsscale{1.10}
\plotone{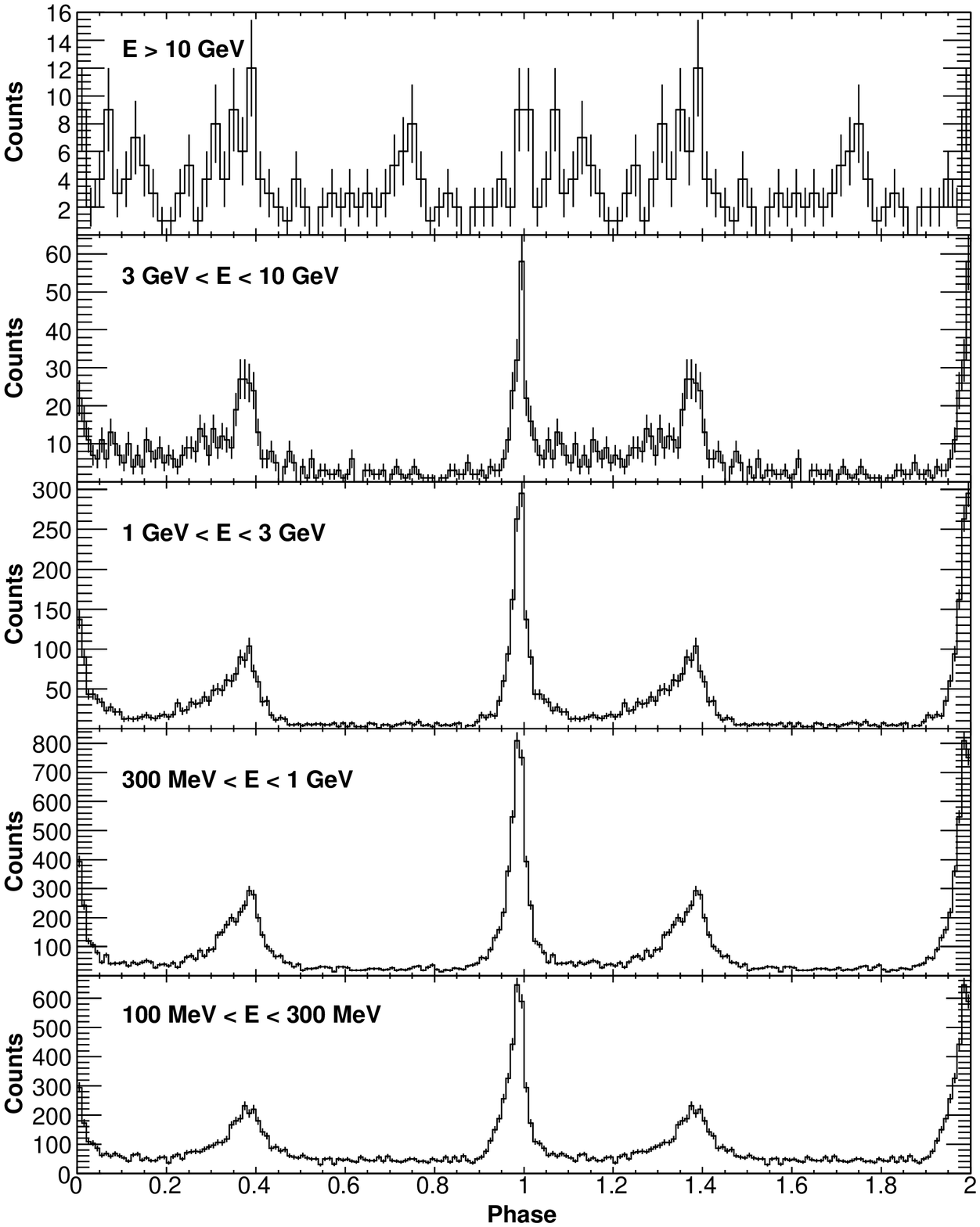}
\caption{\label{phasos_six}\emph{Fermi} light curves for the Crab Pulsar in different energy bands within an energy-dependent circular region, as described in Section~\ref{phasos}. The light curve profile is binned to 0.01 of pulsar phase, except above 10 GeV, which is binned to 0.02 in phase. Two cycles are shown.}
\end{figure*}

The phase 0 is taken at the maximum of the main radio peak observed at 1.4 GHz, as seen in Figures~\ref{phaso_general} and \ref{phaso_multilambda} (a). Considering all events between 100 MeV and 300 GeV, two clear peaks P1 and P2 can be seen at phase $\phi_1$~=~0.9915~$\pm$~0.0005 and $\phi_2$~=~0.3894~$\pm$~0.0022, respectively. Hence, the peaks are separated by $\delta\phi$~=~0.398~$\pm$~0.003. P1 and P2 are asymmetric. Their shapes can be well modeled by two half-Lorentzian functions (with different widths for the leading and trailing sides). The first peak presents rising and falling edges of half-widths 0.045~$\pm$~0.002 and 0.023~$\pm$~0.001 respectively. P2 shows a slow rise and a steeper fall. The rising and falling edges of P2 have Lorentzian half-widths of 0.115~$\pm$~0.015 and 0.045~$\pm$~0.008 respectively. Hence, the $\gamma$-ray first peak leads the radio main pulse by phase 0.0085~$\pm$~0.0005, as shown in Figure~\ref{phaso_general},  where the radio profile (red line) is overlaid for comparison. 

The second $\gamma$-ray peak leads the second 1.4 GHz radio pulse (interpulse) by 0.0143~$\pm$~0.0022 in phase. The peak separation is slightly wider at 1.4 GHz than in $\gamma$-rays.

An error in these $\gamma$-radio delays can also arise from the measurement of the dispersion measure and its derivative. Following \cite{Manchester and Taylor 1977}, the error on the dispersion delay in the propagation of a signal at a frequency $f$ through the interstellar medium is:
\begin{eqnarray}
\Delta(\Delta t) = - \frac{\Delta DM}{K f^2}
\end{eqnarray}
where $\Delta DM$ takes into account the error on the measurement of DM and its derivative, and K = 2.410 $\times$~10$^{-4}$ MHz$^{-2}$~cm$^{-3}$~pc~s$^{-1}$ is the dispersion constant. This yields a formal uncertainty of 1.4~$\mu$s, which is significantly smaller than the 21.1~$\mu$s accuracy of the overall timing solution, and therefore leads to an error of 0.0006 in phase on the $\gamma$-radio delay.

The presence of a radio feature referred to as Low Frequency Component (LFC) by \cite{Moffett and Hankins 1996} can be noticed, at phase 0.896~$\pm$~0.001 on the radio light curve obtained at 1.4~GHz as seen in Figures~\ref{phaso_general} and \ref{phaso_multilambda} (a). This peak is assumed to be near the closest approach of the magnetic axis. The first $\gamma$-ray peak lags the LFC by 0.095~$\pm$~0.002 in phase.

Figure~\ref{phasos_six} shows the light curves in 5 energy bands, covering the 100 MeV -- 300 GeV interval while Table~\ref{table_phaso} reports the evolution of the positions of the peak maxima ($\phi_1$ and $\phi_2$ for P1 and P2 respectively) and their half-widths (HW), for the energy bins between 100~MeV and 10~GeV. The photon number counts above 10 GeV were not sufficient to fit the peak profiles. The phases of the first (P1) and second (P2) peaks do not show any significant shift with energy. Both become narrower when the energy increases, showing in particular a steepening in the P2 falling edge. 

Table~\ref{table_phaso} also presents the energy dependence of the relative weight of the two peaks. The diffuse and nebular background photon density has been first estimated in the 0.52 -- 0.87 phase interval, then renormalized and subtracted so as to determine the number of pulsed photons in both peaks. P1 and P2 are here defined in the 0.87 -- 1.07 and 0.27 -- 0.47 phase intervals respectively. As for the Vela pulsar \citep{Abdo et al. 2009a}, the ratio P1/P2 decreases with increasing energy, especially above a few GeV.

We define the off-pulse window as the 0.52 -- 0.87 phase range, due to the bright emission of the pulsar in the rest of the phase. In the light curve above 10 GeV, we can notice an enhancement indicating a potential third peak at phase $\sim$~0.74, coincident with the radio peak observed between 4.7 and 8.4~GHz and referred to as High Frequency Component 2 (HFC2) by \cite{Moffett and Hankins 1996}. The excess above the background level (estimated at 2.10 counts per bin, with a bin width of 0.02 in phase) is 13.8 photons in the off-pulse interval. The statistical significance of this third peak, 2.3~$\sigma$, is therefore too low to claim a definite detection and a third peak will not be considered separately in the analysis of the Crab Nebula.

\begin{figure*}
\epsscale{1.15}
\plotone{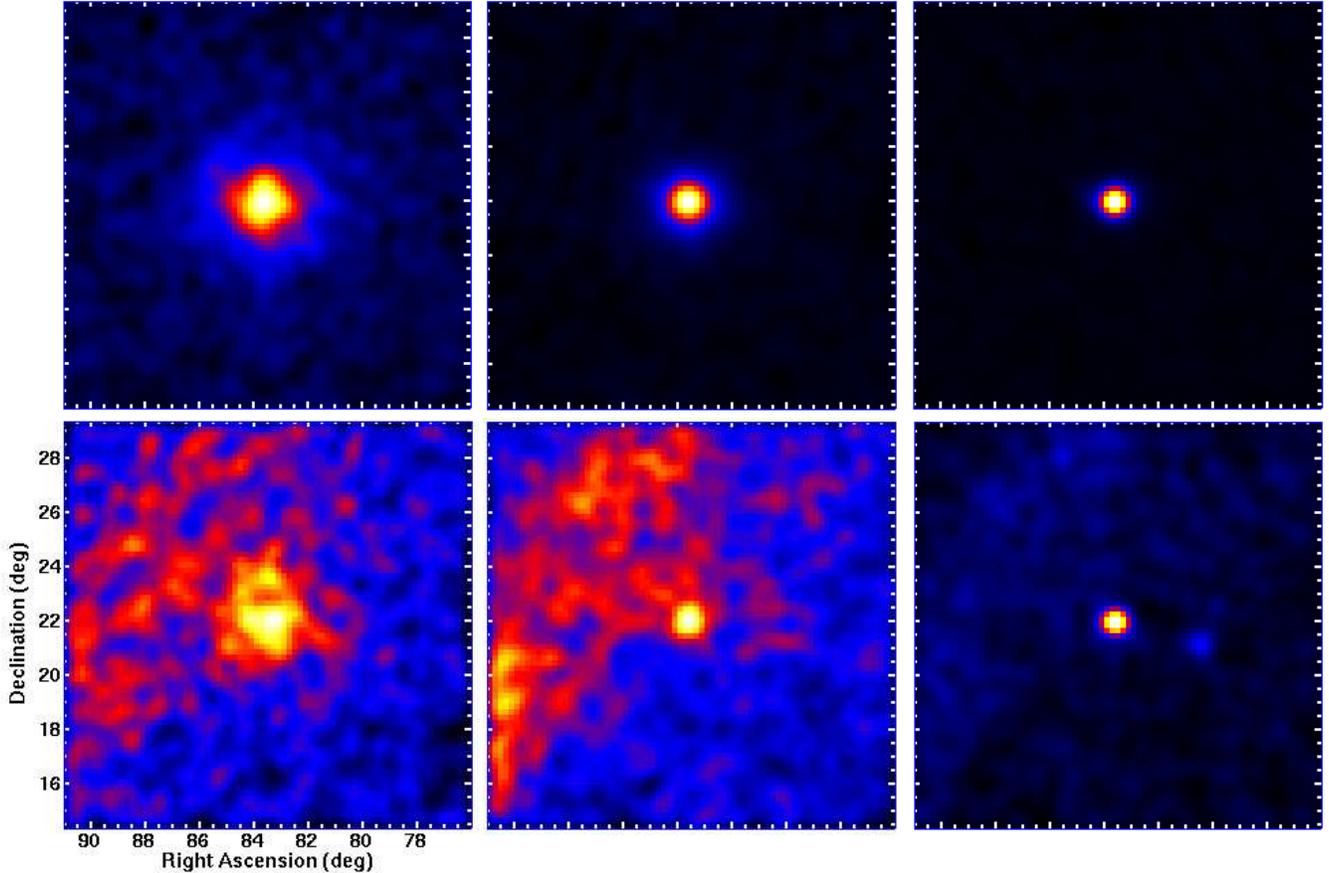}
\caption{\label{CMAP}Counts maps (arbitrary units) presenting the pulsed (top row) and nebular (bottom row) emission, in three energy bands. Each panel spans 15$^{\circ}$ $\times$ 15$^{\circ}$ in equatorial coordinates and is centered on the pulsar radio position. \emph{Left}: 100~MeV~$<$~E~$<$~300~MeV; \emph{Middle}: 300~MeV~$<$~E~$<$~1~GeV; \emph{Right}: E~$>$~1~GeV.}
\end{figure*}

Figure~\ref{CMAP} shows the counts maps of pulsed and nebular emission in a 15$^{\circ}$~$\times$~15$^{\circ}$ region centered on the pulsar radio position, for different energy bands. The nebular emission seen in the off-pulse window has been renormalized to the total phase (bottom row) and subtracted from the whole phase emission, to obtain the maps presenting the pulsed emission only (top row). The positions of the pulsar and nebula are coincident to within our angular resolution and the nebula appears as a point-like source. While the pulsed emission dominates in the on-pulse window, the nebula stands out in the off-pulse interval from the emission of the diffuse background at high energy only.

\begin{table*}[ht]
\begin{center}
\caption{\label{table_phaso}Detailed parameters of the Crab Pulsar light curve. }
\begin{tabular}{cccccccc}
\hline\hline
Energy interval  & $\phi_1$	       & HW$_{1}^{a}$ & HW$_{1}^{b}$   & $\phi_2$	     & HW$_{2}^{a}$	 & HW$_{2}^{b}$ 				     & P1/P2 ratio   \\
(GeV)	       &($\times 10^{-2}$)     & ($\times 10^{-2}$)	       & ($\times 10^{-2}$)	     & ($\times 10^{-2}$)	     &  ($\times 10^{-2}$)	      & ($\times 10^{-2}$)	   &	     \\
\hline
0.1 -- 300     & 99.2 $\pm$ 0.1$^{c}$  & 4.5 $\pm$ 0.2$^{c}$   & 2.3 $\pm$ 0.1$^{c}$	 &38.9 $\pm$ 0.2$^{c}$   &11.5 $\pm$ 1.5$^{c}$   &4.5 $\pm$ 0.7$^{c}$				     & 1.60 $\pm$ 0.06 \\
0.1 -- 0.3	& 99.2 $\pm$ 0.1      & 6.0 $\pm$ 0.4	& 2.3 $\pm$ 0.2 			 &38.3 $\pm$ 0.8     &8.4 $\pm$ 2.2  &8.1 $\pm$ 3.7					     & 1.73 $\pm$ 0.12 \\
0.3 -- 1.0	& 99.1 $\pm$ 0.1      & 4.3 $\pm$ 0.2	& 2.7 $\pm$ 0.2 			 &39.3 $\pm$ 0.3     &13.6 $\pm$ 1.7 &3.6 $\pm$ 0.9					     & 1.60 $\pm$ 0.08 \\
1.0 -- 3.0	& 99.2 $\pm$ 0.1      & 3.5 $\pm$ 0.3	& 2.3 $\pm$ 0.3 			 &38.2 $\pm$ 0.5     &8.5 $\pm$ 2.8  &6.0 $\pm$ 1.9					     & 1.49 $\pm$ 0.12 \\
3.0 -- 10.0	& 99.5 $\pm$ 0.2      & 2.6 $\pm$ 0.5	& 1.6 $\pm$ 0.6 			 &38.9 $\pm$ 0.6     &5.3 $\pm$ 1.9  &2.0 $\pm$ 1.4					     & 0.95 $\pm$ 0.20 \\
\hline  
\multicolumn{8}{l}{$a$, $b$: These half-width (HW) parameters were obtained considering two half-Lorentzian distributions, for the rising and}\\
\multicolumn{8}{l}{falling edges of the peaks respectively. }\\
\multicolumn{8}{l}{$c$: These parameters were derived from a light curve binned to 0.002 of pulsar phase.}
\end{tabular}
\end{center}
\end{table*}

\subsection{Spectral analysis of the Crab Nebula}
\label{nebula}

The spectral analysis of the $\gamma$-ray emission of the Crab Nebula was performed using a maximum-likelihood method \citep{Mattox et al. 1996} implemented in the \emph{Fermi} Science Support Center science tools as the ``gtlike'' code. This fits a source model to the data along with models for the instrumental, extragalactic and Galactic backgrounds. We used an updated instrument response function, Pass6\_v3, that corrects a pileup effect identified in orbit. We selected photons in the 0.52 -- 0.87 pulse phase window in a 20$^{\circ}$ region around the pulsar radio position. Owing to uncertainties in the instrument performance still under investigation at low energies, only events in the 100 MeV -- 300 GeV energy band are analysed. The Galactic diffuse emission is modeled using GALPROP \citep{Strong et al. 2004a, Strong et al. 2004b} updated to include recent HI and CO surveys, more accurate decomposition into Galactocentric rings, and many other improvements. The GALPROP run designation for our model is 54\_59Xvarh7S. The instrumental background and the extragalactic radiation are described by a single isotropic component with a power-law shape. Sources nearby the Crab with a statistical significance larger than $5 \, \sigma$ are extracted using the analysis procedure described in \cite{Abdo et al. 2009b} but with 6 months of survey data, and taken into account in the study. 

The systematic errors on the spectral parameters are dominated by the uncertainties in the LAT instrument response functions (IRFs). We bracket the energy-dependent effective area with envelopes above and below the nominal curves by linearly connecting differences of (10\%, 5\%, 20\%) at log(E) of (2, 2.75, 4) respectively. This yields the systematic errors cited below.

In parallel to the standard analysis, we have also evaluated the spectrum using an unfolding method based on Bayes' theorem \citep{D'Agostini 1995, Mazziotta 2009}, that allows the reconstruction of the true energy spectrum from the observed one taking into account the dispersions introduced by the instrument response function and without assuming any model for the spectral shape. The results from this analysis are consistent with those from the likelihood analysis.

Using EGRET observations, \cite{de Jager et al. 1996} reported that the inverse Compton component dominates above $\sim$~200 MeV whereas the synchrotron component is more significant at lower energies. Hence, the selected $\gamma$-ray photons should allow the study of both the fall of the synchrotron and the rise of the IC radiation. 

\begin{figure*}[ht!!]
\begin{center}
\begin{minipage}[c]{.95\linewidth}
\begin{center}
\epsscale{1.1}
\plotone{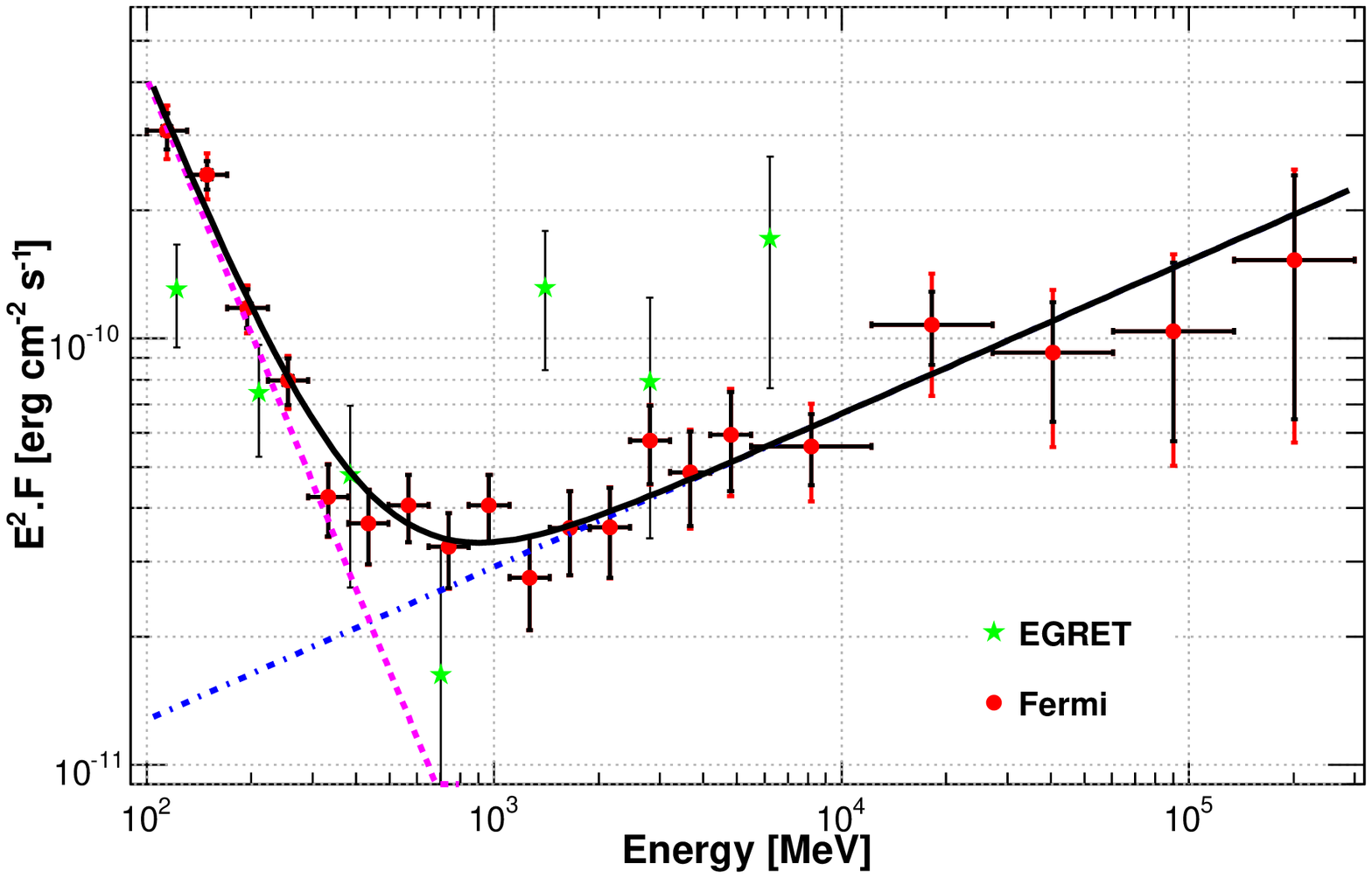}
\caption{\label{SED_neb_8mois}Spectral energy distribution of the Crab Nebula renormalized to the total phase interval. The fit of the synchrotron (purple dashed line) and IC (blue dashed-dotted line) are represented separately with two power-laws. The black curve is the best fit obtained with the sum of these two power-laws.  The LAT spectral points are obtained using the model-independent maximum likelihood method described in Section~\ref{nebula}. The statistical errors are shown in black, while the red lines take into account both the statistical and systematic errors. Horizontal bars delimit the energy intervals. EGRET data points \citep{Kuiper et al. 2001} are shown for comparison (green stars).}
\end{center}
\end{minipage} \vfill
\begin{minipage}[c]{.95\linewidth}
\begin{center}
\epsscale{1.1}
\plotone{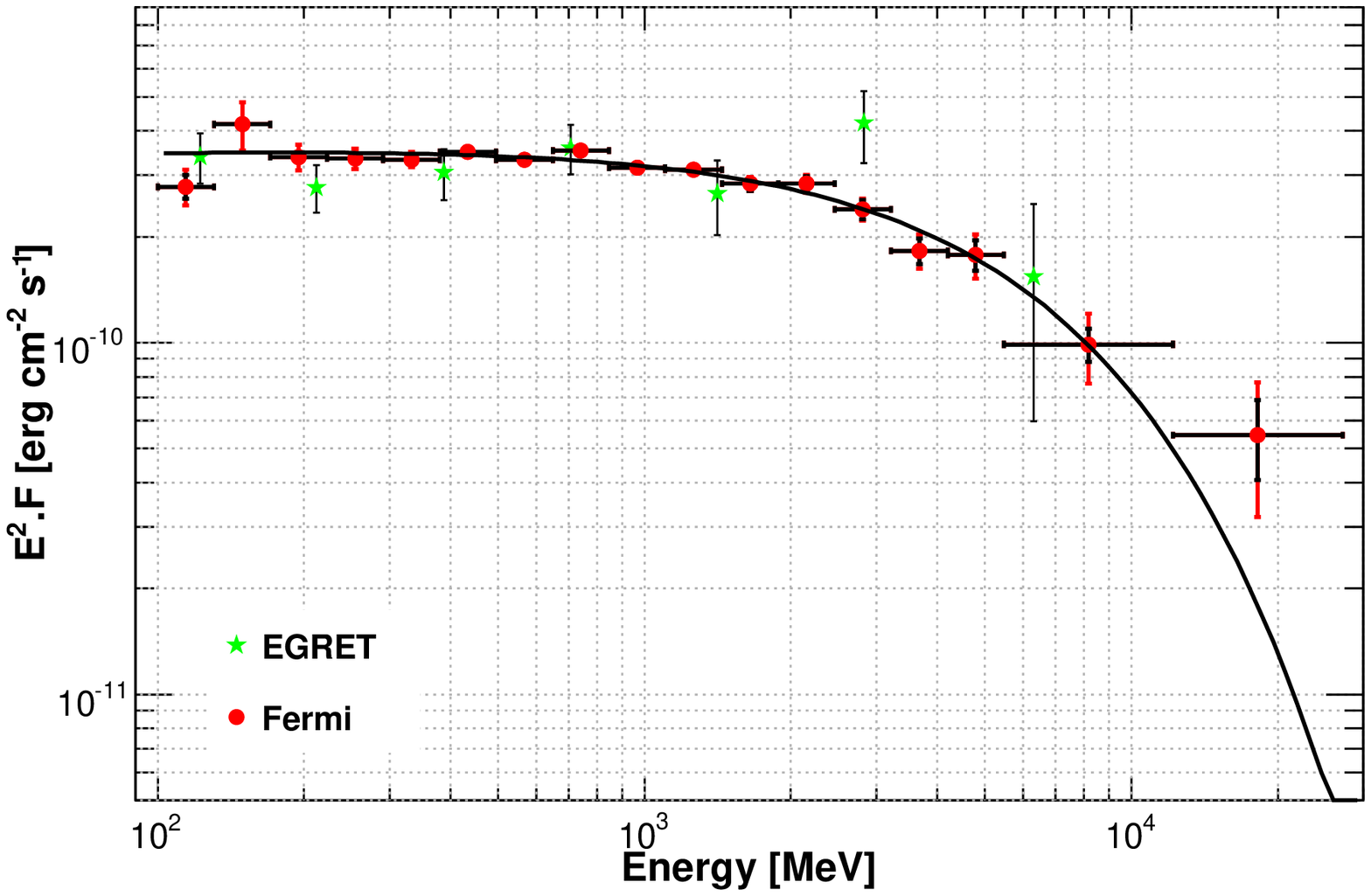}
\caption{\label{SED_pulsar_8mois}Spectral energy distribution of the Crab Pulsar averaged over the whole pulse period. The black curve represents the best fit model, obtained with a power-law with an exponential cut-off. The LAT spectral points (cf. Figure~\ref{SED_neb_8mois} for the description of the conventions) are obtained using the model-independent maximum likelihood method described in Section~\ref{nebula}. EGRET data points \citep{Kuiper et al. 2001} are shown for comparison (green stars).}
\end{center}
\end{minipage}
\end{center}
\end{figure*}

The spectrum of the Crab Nebula between 100 MeV and 300 GeV is well described by the sum of two power-law spectra. As seen on the spectral energy distribution in Figure~\ref{SED_neb_8mois}, one of the components decreases while the second increases with energy. We identify them as the falling edge of the synchrotron component and the rising edge of the IC component, respectively. The nebular spectrum can be modeled with the following function:
\begin{eqnarray}
\frac{dN}{dE} = N_{sync} (E_{GeV})^{-\Gamma_{sync}} + N_{IC} (E_{GeV})^{-\Gamma_{IC}} \nonumber\\
  \, \rm{cm^{-2} \, s^{-1} MeV^{-1}} 
\end{eqnarray}
where $N_{sync}$~=~(9.1~$\pm$~2.1~$\pm$~0.7)~$\times$~10$^{-13}$ cm$^{-2}$~s$^{-1}$ MeV$^{-1}$, $N_{IC}$~=~(6.4~$\pm$~0.7~$\pm$~0.1)~$\times$~10$^{-12}$ cm$^{-2}$~s$^{-1}$ MeV$^{-1}$ are the prefactors determined on 35\% of the total phase, $\Gamma_{sync}$~=~($3.99~\pm~0.12~\pm~0.08$) and $\Gamma_{IC}$~=~($1.64~\pm~0.05~\pm~0.07$) the spectral indices of the synchrotron and IC components. While the power-law index $\Gamma_{IC}$ for the inverse Compton component provides a measure of the index of the mean electron/positron energy spectrum in the nebula, the synchrotron index $\Gamma_{sync}$ possesses much less physical information, being just an indication of the steepness of the quasi-exponential turnover of the synchrotron component that peaks below the LAT energy window. Adopting a power-law fit to the synchrotron contribution apparent in the 100~--~400 MeV range is therefore a useful mathematical convenience. The corresponding flux above 100 MeV and renormalized to the total phase is (9.8~$\pm$~0.7~$\pm$~1.0)~$\times$~10$^{-7}$ cm$^{-2}$~s$^{-1}$. The first error is statistical, whereas the second is systematic. 
 
Figure~\ref{SED_neb_8mois} shows the spectral energy distribution in E$^2 \frac{dN}{dE}$ of the Crab Nebula renormalized to the total phase. The \emph{Fermi}-LAT spectral points were obtained by dividing the 100 MeV -- 300 GeV range into logarithmically-spaced energy bins and performing a maximum likelihood spectral analysis in each interval, assuming a power-law shape for the source. Above 5.5~GeV, the width of the energy intervals is multiplied by 3 to reduce the statistical uncertainties. These points, providing a model-independent maximum likelihood spectrum, are overlaid with the fitted model described above over the total energy range (black curve). The fit of the synchrotron (purple dashed line) and IC (blue dashed-dotted line) are also represented. The spectral points and the model agree well. Statistical errors (black error bars) and the overall error (red error bars) are plotted for the \emph{Fermi} points. The EGRET spectral points are represented on the same plot. As in the case of the spectrum of the Vela pulsar \citep{Abdo et al. 2009a}, derived using an earlier set of response functions, Pass6\_v1, markedly different from Pass6\_v3 at low energies, the LAT spectral points at high energy indicate a lower flux in comparison to EGRET. However, it can be noticed that the \emph{Fermi} flux is higher than the EGRET flux, in the low energy band dominated by synchrotron radiation.

\cite{de Jager et al. 1996} found evidence in the EGRET data that the Crab synchrotron cut-off energy varied on time scales of the order of a year. We do not see significant variation in either the synchrotron or inverse Compton components in our more limited data span on time scales of one, two, or four months. \textbf{As shown in Figure~\ref{SED_neb_8mois}, a difference in flux is observed between EGRET and \emph{Fermi}-LAT in the energy band dominated by synchrotron radiation as well as at higher energies (above 1 GeV). Even if variability in the synchrotron tail could be expected between EGRET and LAT, the lifetimes of the electrons producing gamma-rays via inverse Compton scattering are comparable to the remnant age, implying that the IC component should be steady in time. For these reasons, the flux change seen in the synchrotron component between EGRET and \emph{Fermi}-LAT cannot be considered as significant.}

The photon counts at high energy are too few for a significant cut-off or break to be seen in the flux distribution of the IC component. No cut-off or break energy can be determined at low energy for the synchrotron component using the LAT data only.  
\subsection{Spectral analysis of the pulsed emission}
\label{pulsar}
Photons from both on- and off-pulse intervals are now considered to analyze the pulsed emission. The spectral parameters of the Crab Nebula mentioned in the previous section have been renormalized to match the total phase interval and fixed to perform the spectral analysis of the Crab Pulsar.

After testing different functional forms to describe the spectrum of the pulsar, we found the best fit to be given by an exponential cut-off power-law shape:
\begin{eqnarray}
\frac{dN}{dE} = N_o (E_{GeV})^{-\Gamma} e^{-E/E_{c}} \, \rm{cm^{-2} \, s^{-1} MeV^{-1}}
\end{eqnarray}
where $N_o$~=~($2.36~\pm~0.06~\pm~0.15$)~$\times$~10$^{-10}$ cm$^{-2}$~s$^{-1}$~MeV$^{-1}$ is the prefactor, $\Gamma$~=~($1.97~\pm~0.02$ $\pm~0.06$) the spectral index and $E_{c}$~=~($5.8~\pm~0.5~\pm~1.2$) GeV the cut-off energy of the distribution. The integral flux above 100 MeV is equal to $(2.09~\pm~0.03~\pm 0.18$)~$\times$~10$^{-6}$ cm$^{-2}$~s$^{-1}$. These results are consistent with the pulsed spectrum derived from the unfolding analysis.

Figure~\ref{SED_pulsar_8mois} shows the spectral energy distribution of the Crab Pulsar over the whole pulse period compared to EGRET spectral points. Results of both experiments agree well in the 100~MeV~--~8~GeV energy range, i.e. even at low energies, where such consistency is not observed in the spectrum of the Crab Nebula. The larger energy band covered by the LAT and its better sensitivity allows us to determine the cut-off energy of the spectrum, which was not possible with EGRET.

We also attempted to fit the data using a power-law with a generalized cut-off of the form $e^{-(E/E_c)^b}$ and found $b=(0.89~\pm~0.12~\pm~0.28$) with a likelihood value which is not significantly better than that obtained in the case of a simple exponential $b=1$ cut-off. We compute the probability of incorrect rejection of other spectral shapes using the likelihood ratio test. For instance, if only statistical errors are included, the power-law and hyper-exponential $b=2$ hypothesis shapes are rejected at a level of 10.7~$\sigma$ and 4.9~$\sigma$ respectively.

\begin{table*}[ht!!]
\begin{center}
\caption{\label{table_phase_resolved}Phase interval definitions and corresponding spectral parameters. }
\begin{tabular}{cccccc}
\hline\hline
$\phi_{min}$	    & $\phi_{max}$	   & Flux$^{a}$ 	& Spectral index	 & Cut-off energy (GeV) &  $\chi^{2,b}$ \\
\hline
0.870	    & 0.955	   & 21.0 $\pm$ 1.1  $\pm$ 3.3         & 2.03 $\pm$ 0.12 $\pm$ 0.20	   & 1.7 $\pm$ 0.5 $\pm$ 0.3	    & 5.1$\sigma$\\
0.955	    & 0.971	   & 106.1 $\pm$ 6.1  $\pm$ 18.1	& 2.05 $\pm$ 0.07 $\pm$ 0.22	    & 2.5 $\pm$ 0.6 $\pm$ 0.6	     & 6.0$\sigma$\\
0.971	    & 0.981	   & 177.3 $\pm$ 9.2  $\pm$ 14.0	& 1.97 $\pm$ 0.07 $\pm$ 0.07	    & 2.8 $\pm$ 0.6 $\pm$ 0.2	     & 7.2$\sigma$\\
0.981	    & 0.987	   & 232.8 $\pm$ 14.4  $\pm$ 15.0	 & 1.94 $\pm$ 0.08 $\pm$ 0.05	     & 2.7 $\pm$ 0.7 $\pm$ 0.2        & 6.4$\sigma$\\
0.987	    & 0.993	   & 264.0 $\pm$ 11.5  $\pm$ 11.7	 & 1.93 $\pm$ 0.06 $\pm$ 0.04	     & 4.3 $\pm$ 1.0 $\pm$ 0.4        & 6.9$\sigma$\\
0.993	    & 1.000	   & 205.0 $\pm$ 7.5  $\pm$ 27.1	& 1.90 $\pm$ 0.05 $\pm$ 0.11	    & 5.5 $\pm$ 1.3 $\pm$ 2.0	     & 6.6$\sigma$\\
0.000	    & 0.016	   & 94.8 $\pm$ 3.9  $\pm$ 5.0         & 1.84 $\pm$ 0.08 $\pm$ 0.03	   & 3.1 $\pm$ 0.9 $\pm$ 0.2	    & 6.9$\sigma$\\
0.016	    & 0.098	   & 15.3 $\pm$ 0.9  $\pm$ 2.7         & 1.74 $\pm$ 0.07 $\pm$ 0.20	   & 6.3 $\pm$ 1.8 $\pm$ 2.5	    & 5.9$\sigma$\\
0.098	    & 0.286	   & 5.8 $\pm$ 0.4  $\pm$ 0.4	      & 1.49 $\pm$ 0.09 $\pm$ 0.05	  & 5.5 $\pm$ 1.3 $\pm$ 0.7	   & 7.9$\sigma$\\
0.286	    & 0.338	   & 25.5 $\pm$ 1.3  $\pm$ 3.1         & 1.72 $\pm$ 0.08 $\pm$ 0.14	   & 3.6 $\pm$ 0.8 $\pm$ 0.9	    & 6.8$\sigma$\\
0.338	    & 0.366	   & 52.3 $\pm$ 2.2  $\pm$ 2.5         & 1.94 $\pm$ 0.06 $\pm$ 0.04	   & 6.2 $\pm$ 1.8 $\pm$ 0.3	    & 5.0$\sigma$\\
0.366	    & 0.386	   & 70.4 $\pm$ 2.8  $\pm$ 8.0         & 1.92 $\pm$ 0.05 $\pm$ 0.04	   & 6.8 $\pm$ 1.8 $\pm$ 0.7	    & 5.8$\sigma$\\
0.386	    & 0.410	   & 65.8 $\pm$ 2.7  $\pm$ 13.3        & 2.04 $\pm$ 0.06 $\pm$ 0.16	   & 10.0 $\pm$ 4.8 $\pm$ 11.6  & 3.3$\sigma$\\
0.410	    & 0.520	   & 13.7 $\pm$ 0.8  $\pm$ 1.8         & 2.28 $\pm$ 0.08 $\pm$ 0.10	   & 7.3 $\pm$ 4.8 $\pm$ 2.4	    & 2.3$\sigma$\\
\hline
\multicolumn{6}{l}{$^a$: in units of $10^{-7}$ photons~cm$^{-2}$~s$^{-1}$ and divided by the width of the phase interval. }\\
\multicolumn{6}{l}{$^b$: this value characterizes, for each phase interval, the improvement obtained by using an exponential }\\
\multicolumn{6}{l}{cut-off power-lawinstead of a pure power-law shape, to describe the pulsed spectrum.}
\end{tabular}
\end{center}
\end{table*} 

\subsection{Phase-resolved spectral analysis of the Crab Pulsar}\label{phaseresolved}

The large number of photons detected from the Crab allows a detailed phase-resolved spectroscopic study of its emission. Therefore, the pulse profile is divided in several intervals. The phase bins are chosen so as to contain $\sim$~1000 pulsed photons in the energy-dependent region defined in Section~\ref{phasos}. A maximum-likelihood spectral analysis is performed in each pulse phase interval, assuming a power-law and an exponential cut-off power-law shape to describe the pulsed emission. 

The definition of the phase intervals is given in Table~\ref{table_phase_resolved} along with the spectral results. The last column lists the significance of the improvement obtained when using an exponential cut-off power-law instead of a pure power-law, in terms of $\chi^2$, if only statistical errors are included. 

The corresponding spectral energy distributions are presented in Figure~\ref{resultats_phase_resolved}, where the horizontal error bars delimit the energy intervals. 90\% C.L. upper limits were computed when the statistical significance of the energy interval was lower than 3~$\sigma$. Figure~\ref{phase_resolved_parameters} summarizes the phase-dependence of the variation of the spectral parameters, spectral index, cut-off energy and integral flux above 100 MeV. The vertical error bars take into account both the statistical and systematic uncertainties in the spectral parameters, while the horizontal error bars delimit the phase intervals.

One observes a slight steepening of the spectrum with phase in the interval corresponding to P1 (0.955~--~0.098), with averaged values of spectral index and cut-off energy close to $\sim$~1.9 and $\sim$~3~GeV respectively. 

The 0.098~--~0.286 pulse phase interval presents the hardest spectrum with a spectral index of 1.49~$\pm$~0.09~$\pm$~0.05. This result is consistent with the spectrum of the "bridge", as defined in \cite{Fierro et al. 1998}.

The spectral indices of the two peaks are the same, within the error bars, but the second peak (0.286~--~0.410) is characterized by a cut-off energy apparently larger than that of P1. This difference is consistent with the decrease of the P1/P2 ratio with the energy, especially above a few GeV, observed in Figure~\ref{phasos_six}.

Finally, the 0.410~--~0.520 phase bin has the softest spectrum of the total pulse phase interval with a spectral index of 2.28~$\pm$~0.08~$\pm$~0.10. This explains the trend seen in the phase histograms: in Figure~\ref{phasos_six}, the right edge of the second peak falls with increasing energy. 

\section{Discussion}
\label{discussion}

\subsection{Synchrotron and inverse Compton emission from the Crab Nebula}

The Crab Nebula is detected across the whole electromagnetic spectrum from radio to very high energy $\gamma$-rays. The total spectral energy distribution of this source is shown in Figure~\ref{SED_nebula}, from soft to very-high energy $\gamma$-rays. The spectral points obtained with the LAT data analysis are also represented (red points).

With a spectral index of $\Gamma_{IC}$~=~($1.64~\pm~0.05~\pm~0.07$), the LAT results on the rising edge of the inverse Compton component are consistent with EGRET ($1.85^{+ 0.65}_{-0.5}$, \cite{de Jager et al. 1996}). As can be observed in Figure~\ref{SED_nebula}, the highest part of the LAT spectrum links up satisfactorily to the lower energy bound of the Cherenkov data points. Using the LAT spectral parameters scaled to the full pulse phase, we obtain a flux at 77 GeV of $(1.18~\pm~0.22~\pm~0.37)~\times 10^{-14}$ cm$^{-2}$~s$^{-1}$~MeV$^{-1}$ which agrees with the MAGIC differential flux at this energy of $(1.14~\pm~0.27~\pm~0.34)~\times~10^{-14}$ cm$^{-2}$~s$^{-1}$~MeV$^{-1}$. The Cherenkov and \emph{Fermi}-LAT data now cover the entire inverse Compton peak, as can be seen in Figure~\ref{SED_nebula}, and a break is expected at $\sim 100$~GeV. Although no significant cut-off is observed in the LAT data with the current statistics, the determination of its position with an increased \emph{Fermi}-LAT data sample would help the calibration of Cherenkov telescopes, as discussed in \cite{Bastieri et al. 2005}.

The inverse Compton scattering of relativistic electrons on the synchrotron, far infrared, and cosmic microwave background radiation fields is considered to be the most probable mechanism for production of $\gamma$-rays above 1 GeV. However, using a sophisticated approach carried out in the framework of the MHD flow of \cite{Kennel and Coroniti 1984}, \cite{Atoyan and Aharonian 1996} have commented on the apparent deficit of GeV photons in their calculations. Taking into account both EGRET and Cherenkov results and assuming a mean magnetic field which reproduces the very high energy spectrum, they proposed that the high $\gamma$-ray flux observed by EGRET in comparison to their model is due to the enhancement of the bremsstrahlung emission from electrons captured in dense filaments. Figure~\ref{SED_nebula} presents the broad-band energy spectrum of the Crab Nebula together with the inverse Compton model predictions from Figure~14 of \cite{Atoyan and Aharonian 1996} for three different values of the mean magnetic field for the nebula. In view of the results obtained with the LAT, modeling the data does not require any additional emission component. The \emph{Fermi}-LAT, in combination with the Cherenkov observations above 100~GeV, are in good agreement with the $\gamma$-ray flux predicted from simple IC scattering when the magnetic field lies between 100~$\mu$G and 200~$\mu$G, i.e. below the canonical equipartition field of the Crab Nebula of 300~$\mu$G. This result is consistent with the estimate of the magnetic field strength B~$\sim$~140~$\mu$G obtained by \cite{Horns and Aharonian 2004}. 

Concerning the low energy part of the nebular spectrum, the LAT spectral points, combined with COMPTEL's (taking into account statistical errors only for the latter), can be fitted with a power-law with an exponential cut-off, following \cite{de Jager et al. 1996}. The cut-off energy is estimated at $E_{c, sync}$~=~(97~$\pm$~12)~MeV. The higher value of this energy compared to that of \cite{de Jager et al. 1996} is due to the larger flux obtained with \emph{Fermi} than by EGRET for the synchrotron component. The fit is represented with a blue dashed curve in Figure~\ref{SED_nebula}.

\subsection{High energy emission from the Crab Pulsar}

The high-quality statistics obtained with the \emph{Fermi}-LAT both on the light curve and the spectrum of the Crab Pulsar, allow a more detailed comparison with theoretical models than previously possible. Currently, there are two classes of models that differ in the location of the emission region. The first comprises polar cap (PC) models which place the emission near the magnetic poles of the neutron star \citep{Daugherty and Harding 1996}. The second class consists of the outer gap (OG) models \citep{Romani 1996}, in which the emission extends between the null charge surface and the light cylinder, and the two-pole caustic (TPC) models \citep{Dyks and Rudak 2003} which might be realized in slot gap (SG) acceleration models \citep{Muslimov and Harding 2004}, in which the emission takes place between the neutron star surface and the light cylinder along the last open field lines. 


%
\begin{table}
\begin{center}
\caption{\label{lead/lag_radio}The radio delay with respect to other frequencies. }
\begin{tabular}{ccc}
\hline\hline
Spectral band	      & Radio delay	  & Reference\\
	       & ($\mu$s)  & \\
\hline
Optical 	&255 $\pm$ 21	   & (1) \\
X-rays         &344 $\pm$ 40	  & (2) \\
Hard X-rays	 &280 $\pm$ 40      & (3) \\
$\gamma$-rays (EGRET)	&241 $\pm$ 29	   & (3) \\
$\gamma$-rays (LAT)   &281~$\pm$~12~$\pm$~21	& (4) \\
\hline
\multicolumn{3}{l}{References: (1): \cite{Oosterbroek et al. 2008}; (2): \cite{Rots et al. 2004};}\\
\multicolumn{3}{l}{(3): \cite{Kuiper et al. 2003}; (4): this paper.}
\end{tabular}
\end{center}
\end{table} 

Observations of the time delay between emission at different wavelengths have been reported previously: Table~\ref{lead/lag_radio} summarizes the delay of the radio main pulse with respect to the first peak seen from optical to high energy $\gamma$-rays. The LAT has a timing accuracy better than 1 $\mu$s \citep{Abdo et al. 2009c} and thus enables an accurate estimation of the absolute positions of the $\gamma$-ray peaks: it was shown in Section~\ref{phasos} that the first $\gamma$-ray peak leads the radio main pulse by 0.0085 $\pm$ 0.0005 $\pm$ 0.0006 in phase, or (281~$\pm$~12~$\pm$~21)~$\mu$s in time. Taking into account the presence of the LFC at phase 0.896~$\pm$~0.002, the first radio peak leads the $\gamma$-ray peak by phase 0.095~$\pm$~0.002 in phase. Observations of the evolution of the peak positions with the energy allow detailed studies of the emission regions in the magnetosphere. The delay of the radio peaks compared to other wavelengths (optical, X- and $\gamma$-rays) gives another constraint in the modeling of the emission processes taking place in pulsar magnetospheres. In particular, \cite{Kuiper et al. 2003} reported that, in the framework of the three-dimensional outer gap model developed by \cite{Cheng et al. 2000}, one can reproduce the delay of the radio main pulse by shifting the production site of the radio emission inwards toward the neutron star relative to that of high energy photons.

In the PC models, $\gamma$-rays created near the neutron star surface interact with the intense magnetic fields resulting in a sharp turnover in the few to 10~GeV energy range, while OG and SG models predict a simple exponential cut-off.
Furthermore, the maximum energy of observed pulsed photons must lie below any $\gamma$-B pair production turnover threshold, providing a lower bound to the altitude of emission. We can use the observed phase-averaged cutoff energy ($\sim$~6 GeV) to estimate a minimum emission height as $r \ge (\epsilon_{max} B_{12}/1.76 \, {\rm GeV})^{2/7}P^{-1/7} R_\ast$, where $\epsilon_{max}$ is the unabsorbed photon energy, $P$ is the spin period and the surface field is $10^{12}B_{12}$\,G \citep{Baring 2004}. Using the parameters of the Crab Pulsar ($P~=~33$~ms, $B_{12}=3.78$), one obtains $r > 3.4 R_{\ast}$ which precludes emission near the stellar surface. Since we see pulsed photons up to almost $\epsilon_{max} \approx 20$ GeV in the two main pulse peaks, emission at these phases must arise at $r > 4.8 R_{\ast}$, with a strict lower bound of $r > 3.7 R_{\ast}$ applying to the choice of 8 GeV, the lower energy in the highest data point window for the phase-resolved spectra in Figure~\ref{resultats_phase_resolved}. A similar lower bound to the emission altitude was recently reported by the MAGIC collaboration using the hyper-exponential cutoff energy observed on the Crab Pulsar spectrum \citep{Aliu et al. 2008}. We should note here that the cut-off energy derived by the MAGIC collaboration for a simple exponential cut-off ($17.7 \pm 2.8 \pm 5.0$)~GeV is higher than the one obtained with the \emph{Fermi}-LAT data, $E_{c}$~=~($5.8~\pm~0.5~\pm~1.5$)~GeV. However, the cut-off energy obtained with the LAT using the softer EGRET spectrum ($\gamma$~=~2.022) as done by MAGIC is within the uncertainties of the MAGIC value.  

To estimate the pulsed high energy $\gamma$-ray efficiency $\eta$ of a pulsar, one needs to know the total luminosity radiated $L_{\gamma}$. It can be estimated using $L_{\gamma} = 4\pi f_{\Omega}F_{obs}D^2$, where $F_{obs}$~=~(1.31~$\pm$~0.02~$\pm$~0.02)$~\times$~10$^{-9}$ erg~cm$^{-2}$~s$^{-1}$ is the observed phase-averaged energy flux over 100 MeV, $D$~=~(2.0~$\pm$~0.2)~kpc is the distance to the pulsar and $f_{\Omega}$ is a correction factor that takes into account the beaming geometry, depending upon the magnetic inclination angle $\alpha$ and the Earth viewing angle $\zeta$ from the rotation axis. For the Crab Pulsar, $\zeta$ is estimated to be (63 $\pm$ 2)$^{\circ}$ from X-ray observations of the Crab Nebula torus \citep{Ng and Romani 2008}. The estimated value of $\alpha$ depends on the emission model used to interpret the data \citep{Watters et al. 2009}. The slot gap or two pole caustic model best reproduces the observed pulse profiles for $\alpha \sim$ 55 -- 60$^{\circ}$, whereas for the outer gap model $\alpha \sim 70^{\circ}$ gives the best result. Optical polarization measurements yield $\alpha$ estimates consistent with these~\citep{Slowikowska et al. 2009}. The corresponding correction factors $f_{\Omega}$ are then equal to 1.1 for the TPC and 1.0 for the OG models. This yields a luminosity of (6.25~$\pm$~0.15~$\pm$~0.15)~$\times$~10$^{35}$ erg~s$^{-1}$ above 100 MeV. This value is consistent with the heuristic luminosity law mentioned in \cite{Arons 1996} and \cite{Watters et al. 2009}, according to which $\eta \propto \dot{E}^{-1/2}$ and verified by several $\gamma$-ray pulsars such as Vela, PSR J2021+3651 (assuming a distance of the order of 2 -- 4 kpc), Geminga, CTA1, etc. For a neutron star moment of inertia of 10$^{45}$ g~cm$^{2}$, the pulsed high energy $\gamma$-ray efficiency $\eta$ can be derived from the luminosity and the spin down power $\dot{E}$: $\eta~=~L_{\gamma}~/~\dot{E}~=~(1.36~\pm~0.03~\pm~0.03) \times~10^{-3}$ above 100 MeV. 

Knowing the value of the Earth viewing angle $\zeta$ of (63 $\pm$ 2)$^{\circ}$, the main peak separation, which is of the order of 40\% of the phase, would be expected to be smaller in the Polar Cap model \citep{Watters et al. 2009}. This mismatch persists even if moderate altitudes are considered: the open field line cone opening angle enlargens to around $12.3^{\circ}$ for the Crab at altitudes of around 6 stellar radii, the minimum bound inferred above from the observed absence of magnetic pair attenuation below 20 GeV. This opening angle is still somewhat too small according to the \cite{Watters et al. 2009} analysis to generate the observed main peak phase separation.

\section{Summary}
\label{summary}

Using 8 months of survey data with the \emph{Fermi} Large Area Telescope and the very precise timing solution provided by the Nan\c{c}ay and Jodrell Bank radio telescopes, we have examined the high energy behaviour of the Crab Pulsar and Nebula. The unprecedented statistics of the data improve our knowledge of these two sources and place new constraints on theoretical models:

1. The $\gamma$-ray profile of the pulsar consists of two main peaks (P1 and P2), very stable in position across the $\gamma$-ray energy band and separated by $\delta\phi$ = 0.398 $\pm$ 0.003 in phase. The ratio P1/P2 decreases with energy, as seen for several other pulsars and especially for Vela. The first $\gamma$-ray pulse leads the radio main pulse by (281 $\pm$ 12 $\pm$~21) $\mu$s.

2. The spectrum of the nebula in the energy range 100 MeV -- 300 GeV is well described by the sum of two power-laws of spectral indices $\Gamma_{sync}$~=~($3.99~\pm~0.12~\pm~0.08$) and $\Gamma_{IC}$~=~($1.64~\pm~0.05~\pm~0.07$) describing the synchrotron and inverse Compton components of the Crab Nebula spectrum respectively. No cut-off energy can be estimated for the synchrotron component using the LAT data only. The IC rising edge studied in the LAT energy range extends nicely up to the energy domain covered by Cherenkov experiments. No significant cut-off at high energy is observed with the current statistics in the LAT energy range. No significant variation in either the synchrotron or Compton components is seen with the current statistics on time scales of one, two, or four months.

3. The phase-averaged $\gamma$-ray spectrum of the Crab Pulsar can be represented by a power-law with an exponential cut-off at $E_{c}$~=~($5.8~\pm~0.5~\pm~1.2$)~GeV. The hyper-exponential cut-off index $b = (0.89~\pm~0.12~\pm~0.28)$ is not significantly favoured with respect to the simple exponential $b = 1$. If only statistical errors are included, $b=2$ is rejected at 4.9~$\sigma$ level. Using the observed cut-off energy to estimate the minimum emission height $r$ of the emission region, one obtains $r > 3.4 R_\ast$ which precludes emission near the stellar surface.

4. The pulsar emission is hardest in the phase interval between the peaks, usually called the "bridge", while the softest components is the falling edge of the second peak. Both peaks present similar spectral indices, while the cut-off of P1 is lower than P2, consistent with the energy-dependence of the pulse profiles and of the ratio P1/P2.

5. Knowing the Earth viewing angle $\zeta~\sim~63^{\circ}$ and the value of the inclination angle $\alpha$ comprised between 55 -- 60$^{\circ}$ for slot gap models and $\sim~70^{\circ}$ for the outer gap, one can estimate a pulsed high energy $\gamma$-ray efficiency of $\sim~0.1$\% for the conversion of the spin-down energy to $\gamma$-ray emission.\\

The \emph{Fermi} LAT Collaboration acknowledges generous ongoing support from a number of agencies and institutes that have supported both the development and the operation of the LAT as well as scientific data analysis.  These include the National Aeronautics and Space Administration and the Department of Energy in the United States, the Commissariat \`a l'Energie Atomique and the Centre National de la Recherche Scientifique / Institut National de Physique Nucl\'eaire et de Physique des Particules in France, the Agenzia Spaziale Italiana and the Istituto Nazionale di Fisica Nucleare in Italy, the Ministry of Education, Culture, Sports, Science and Technology (MEXT), High Energy Accelerator Research Organization (KEK) and Japan Aerospace Exploration Agency (JAXA) in Japan, and the K. A. Wallenberg Foundation, the Swedish Research Council and the Swedish National Space Board in Sweden.\\
Additional support for science analysis during the operations phase is gratefully acknowledged from the Istituto Nazionale di Astrofisica in Italy.\\
The Nan\c{c}ay Radio Observatory is operated by the Paris Observatory, associated with the French Centre National de la Recherche Scientifique (CNRS). The Lovell Telescope is owned and operated by the University of Manchester as part of the Jodrell Bank Centre for Astrophysics with support from the Science and Technology Facilities Council of the United Kingdom.

\begin{figure*}[h!]
\begin{minipage}[c]{.98\linewidth}
\epsscale{.58}
\plotone{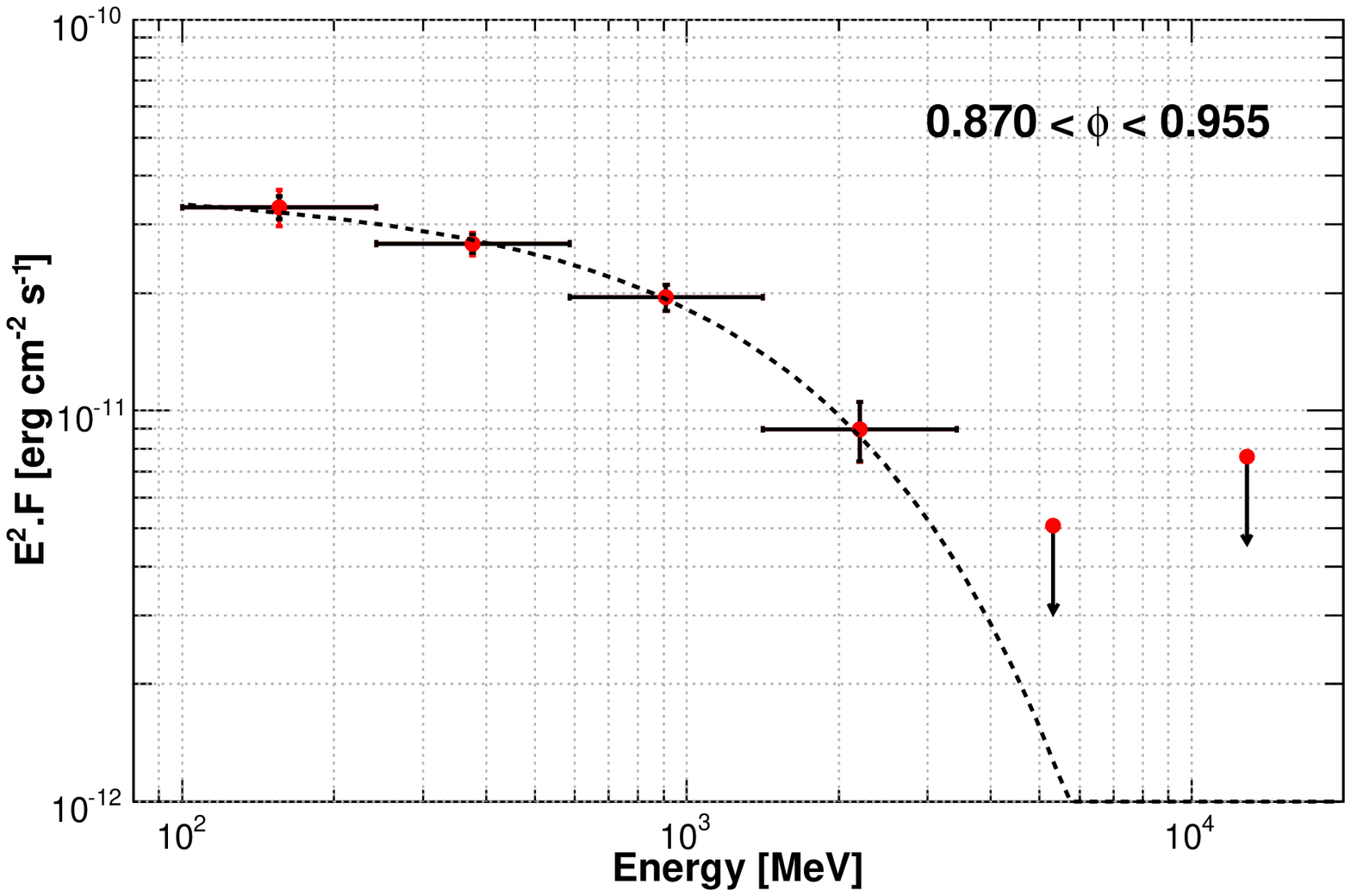}
\plotone{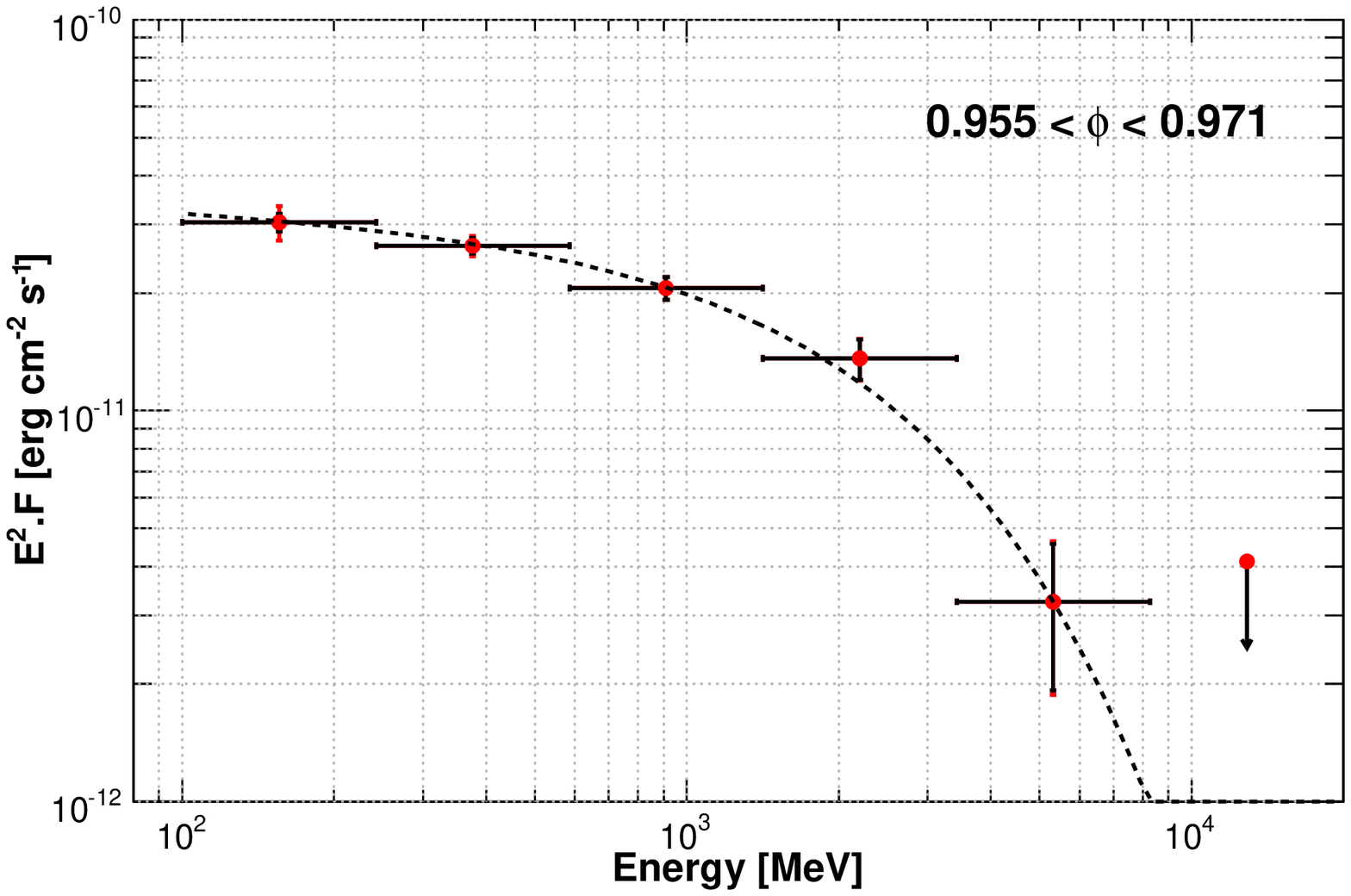}
\end{minipage} \hfill
\begin{minipage}[c]{.98\linewidth}
\epsscale{.58}
\plotone{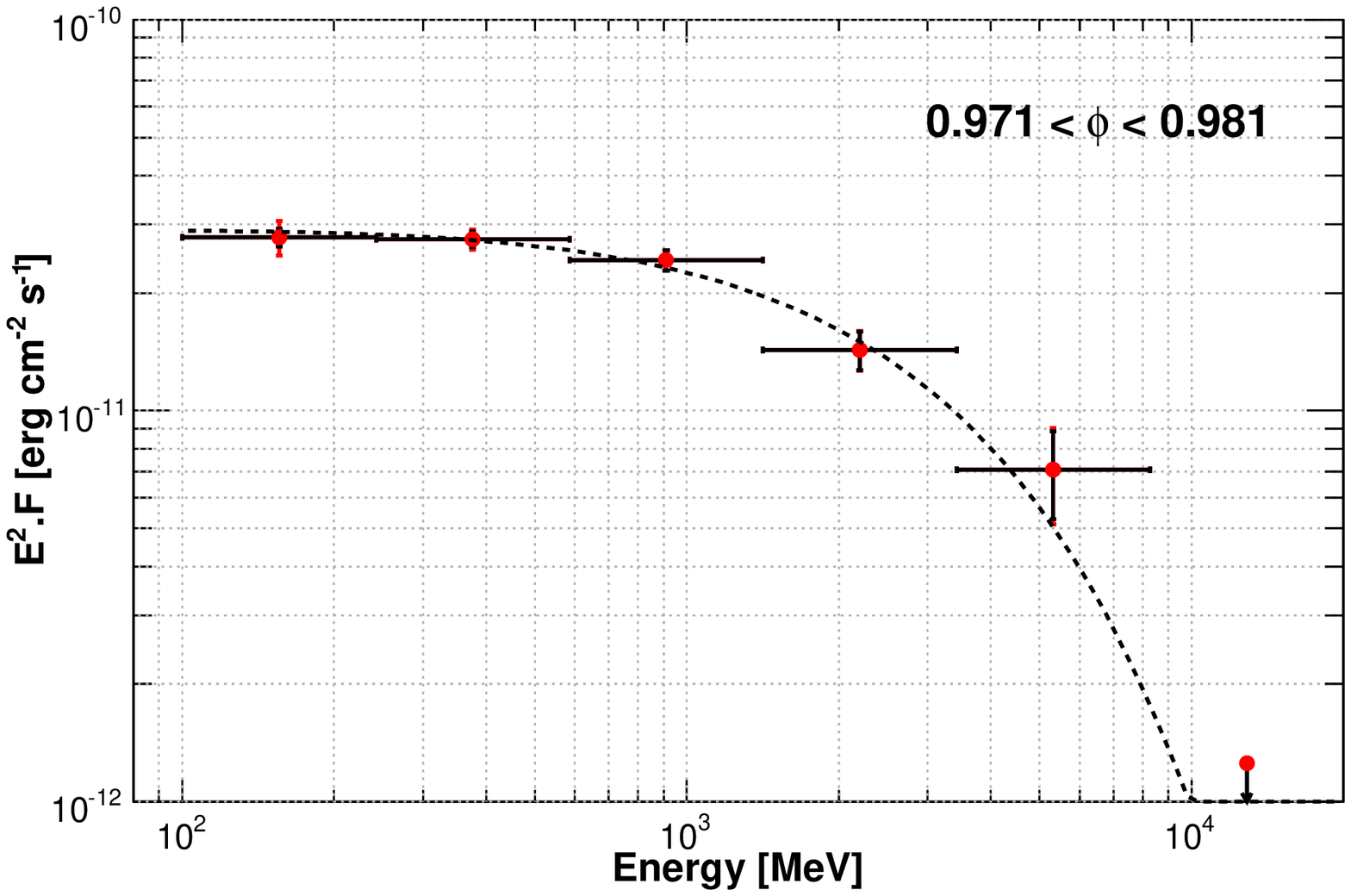}
\plotone{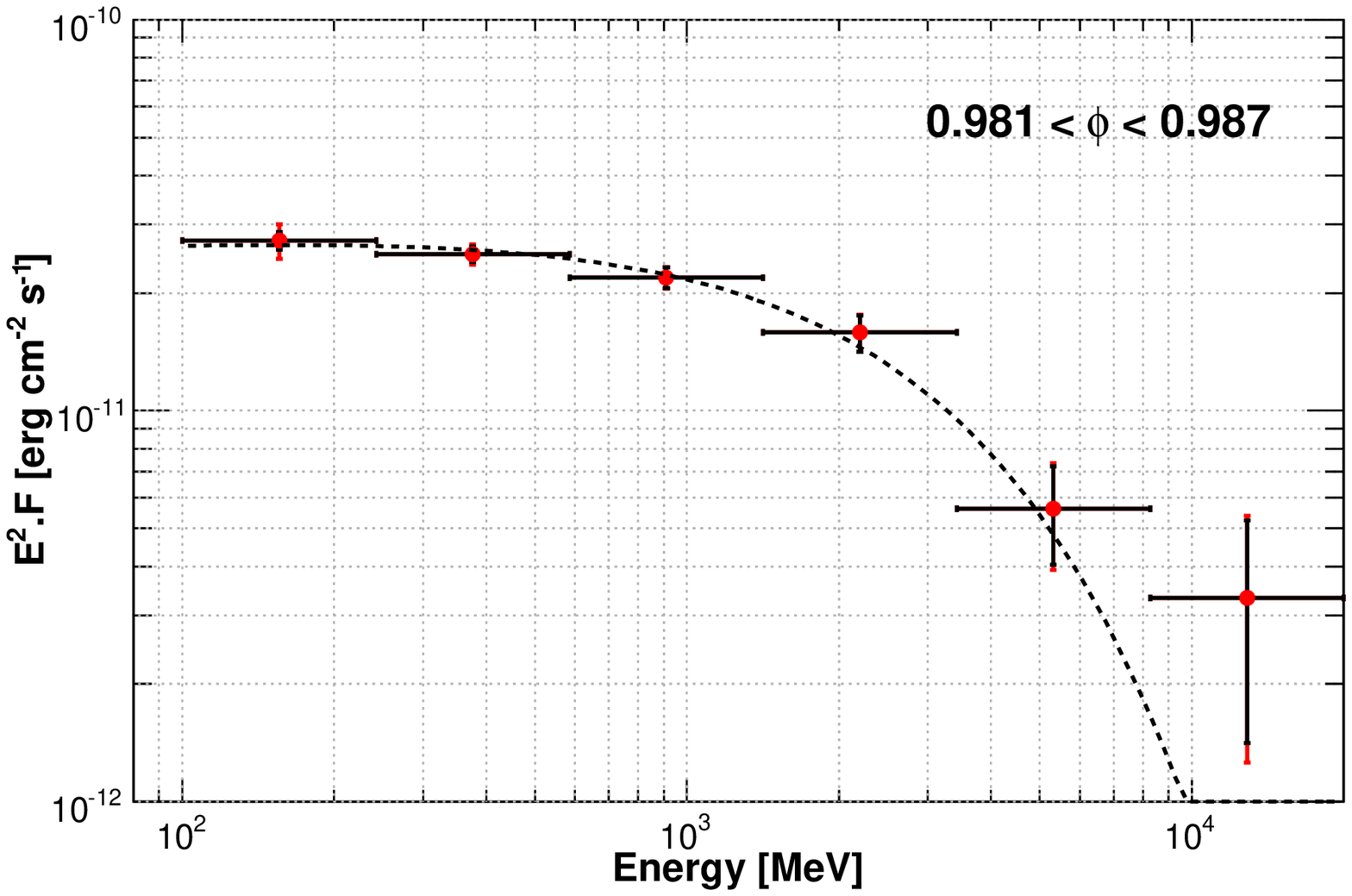}
\end{minipage} \hfill
\begin{minipage}[c]{.98\linewidth}
\epsscale{.58}
\plotone{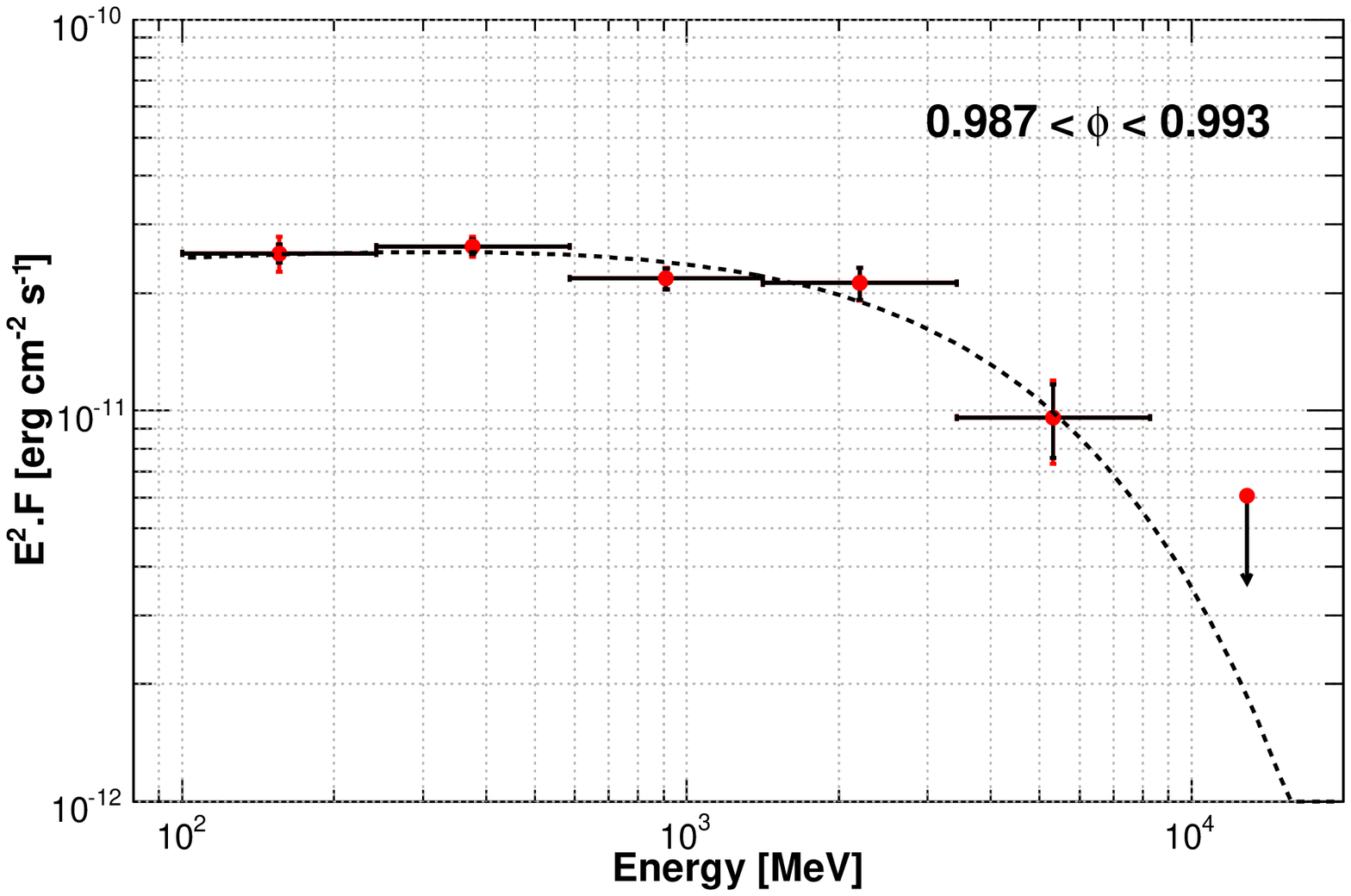}
\plotone{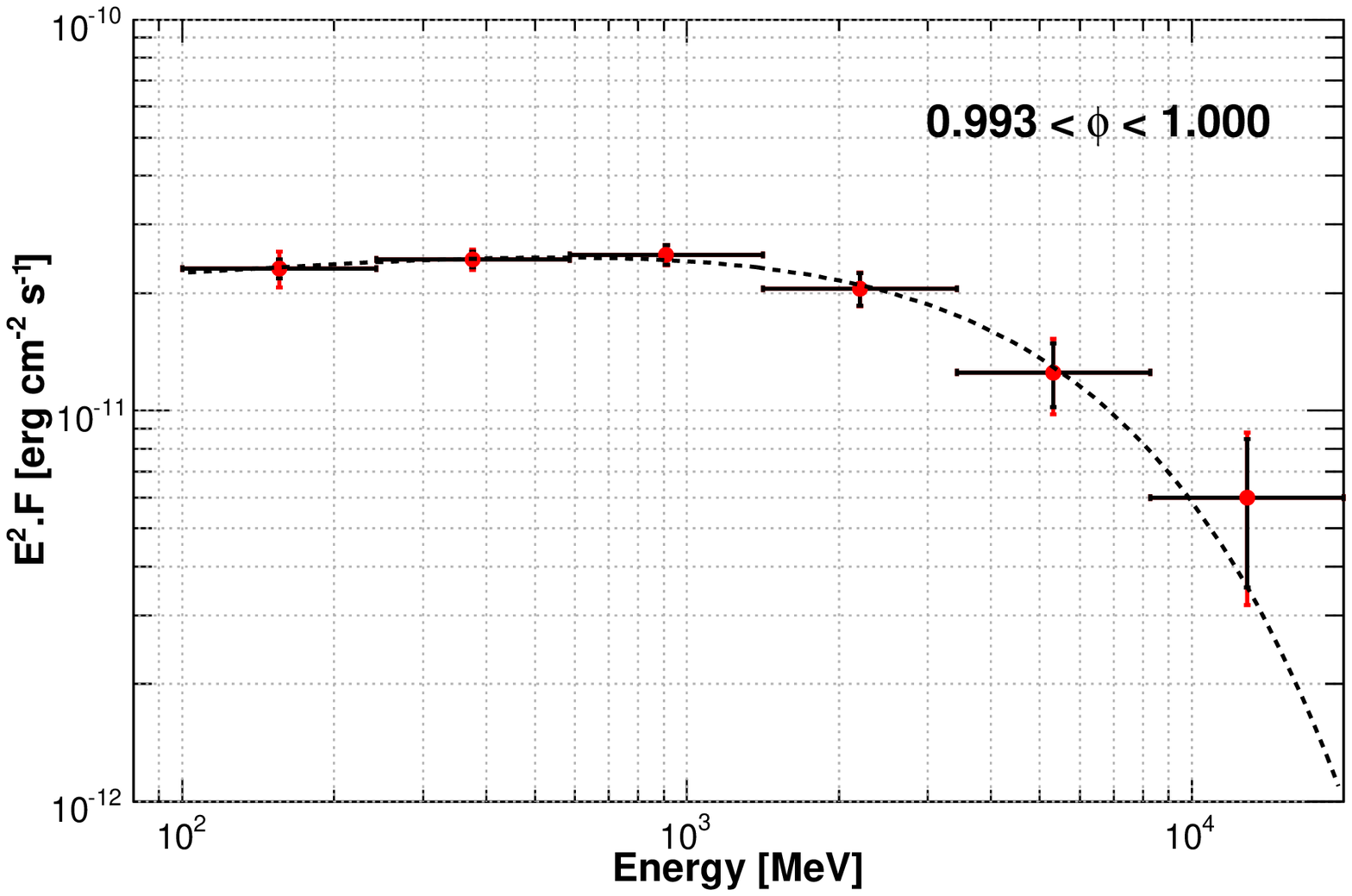}
\end{minipage} \hfill
\begin{minipage}[c]{.98\linewidth}
\epsscale{.58}
\plotone{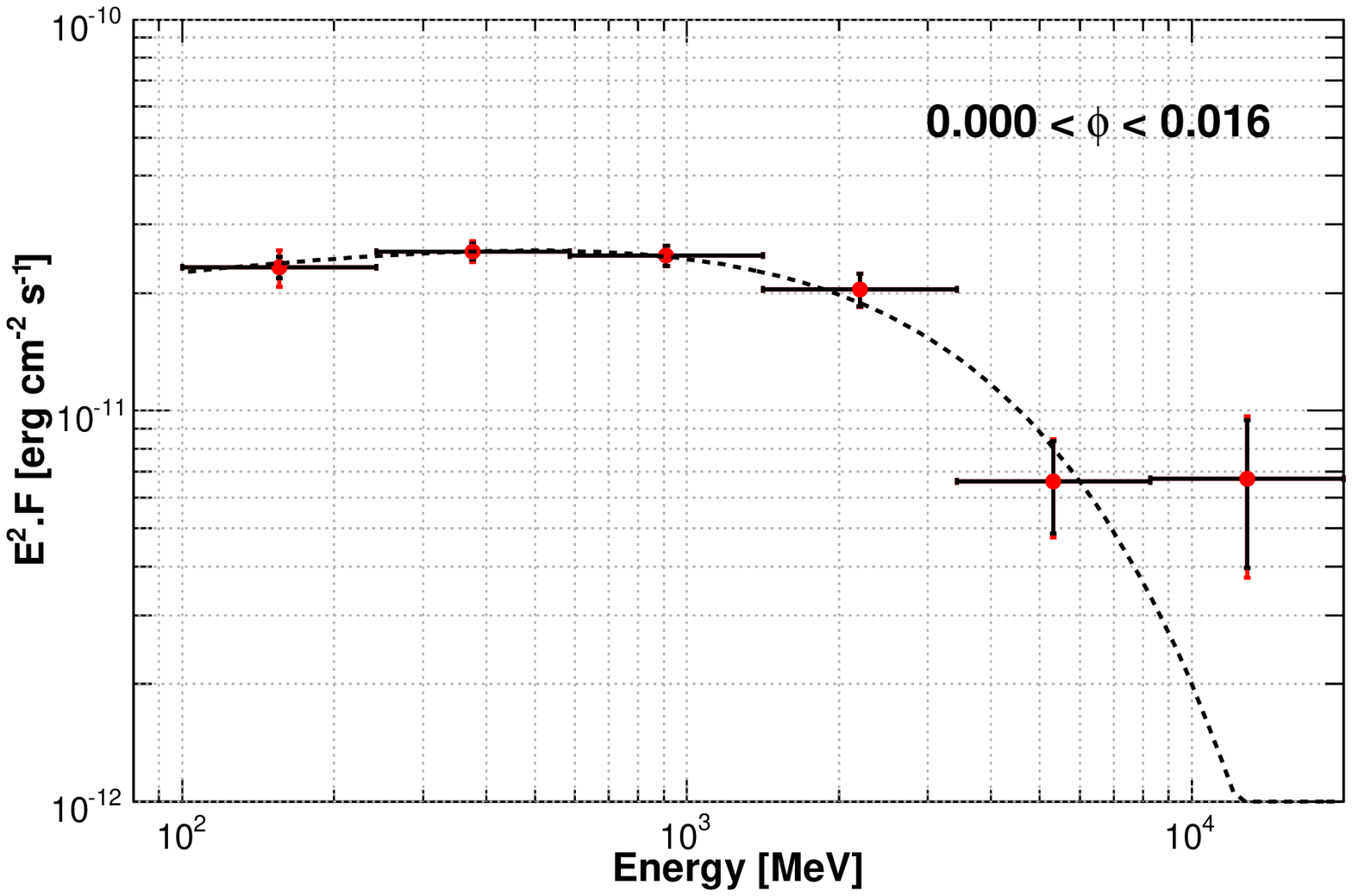}
\plotone{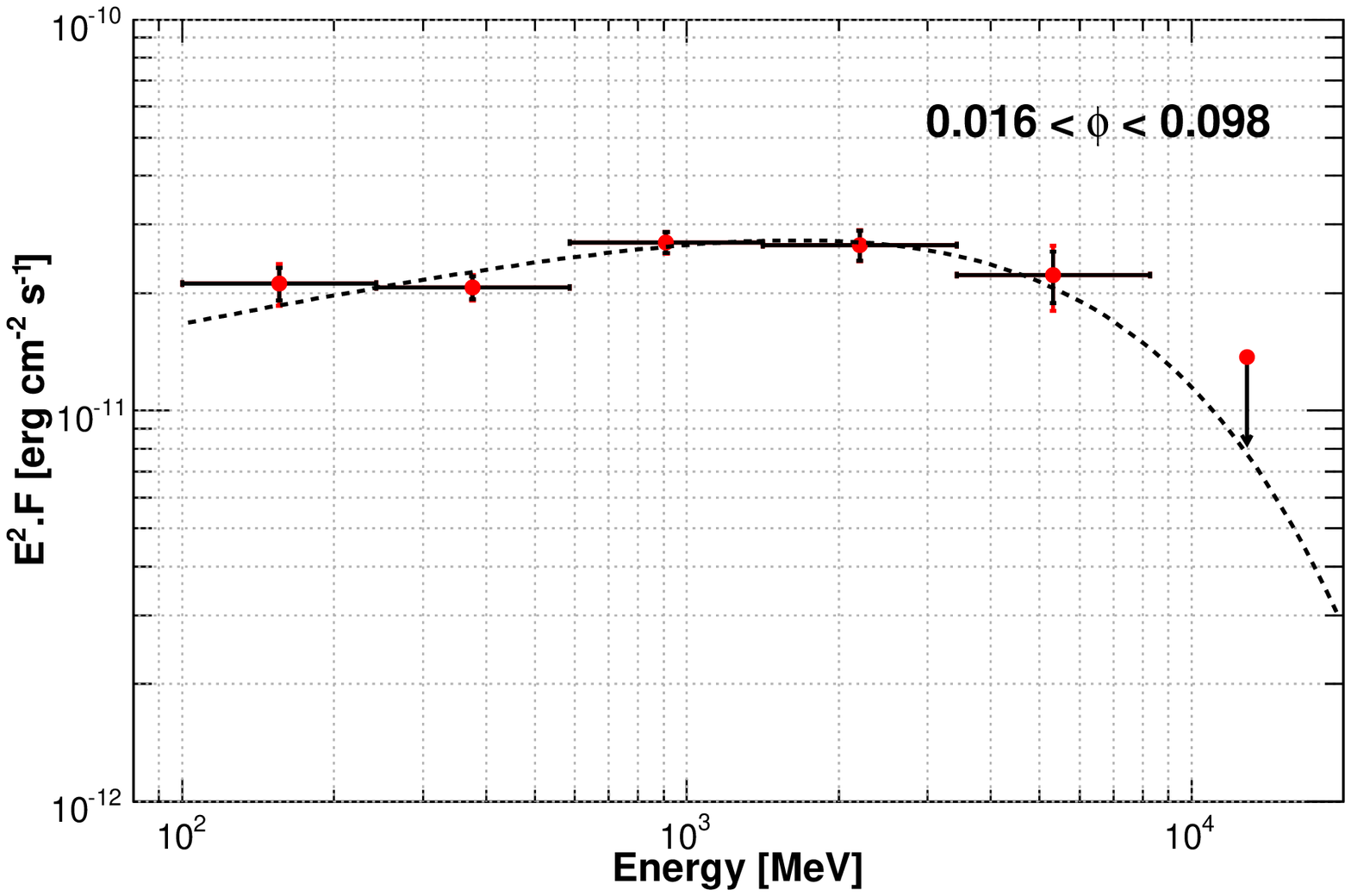}
\end{minipage} \hfill
\end{figure*}

\begin{figure*}[h!]
\begin{minipage}[c]{.98\linewidth}
\epsscale{.58}
\plotone{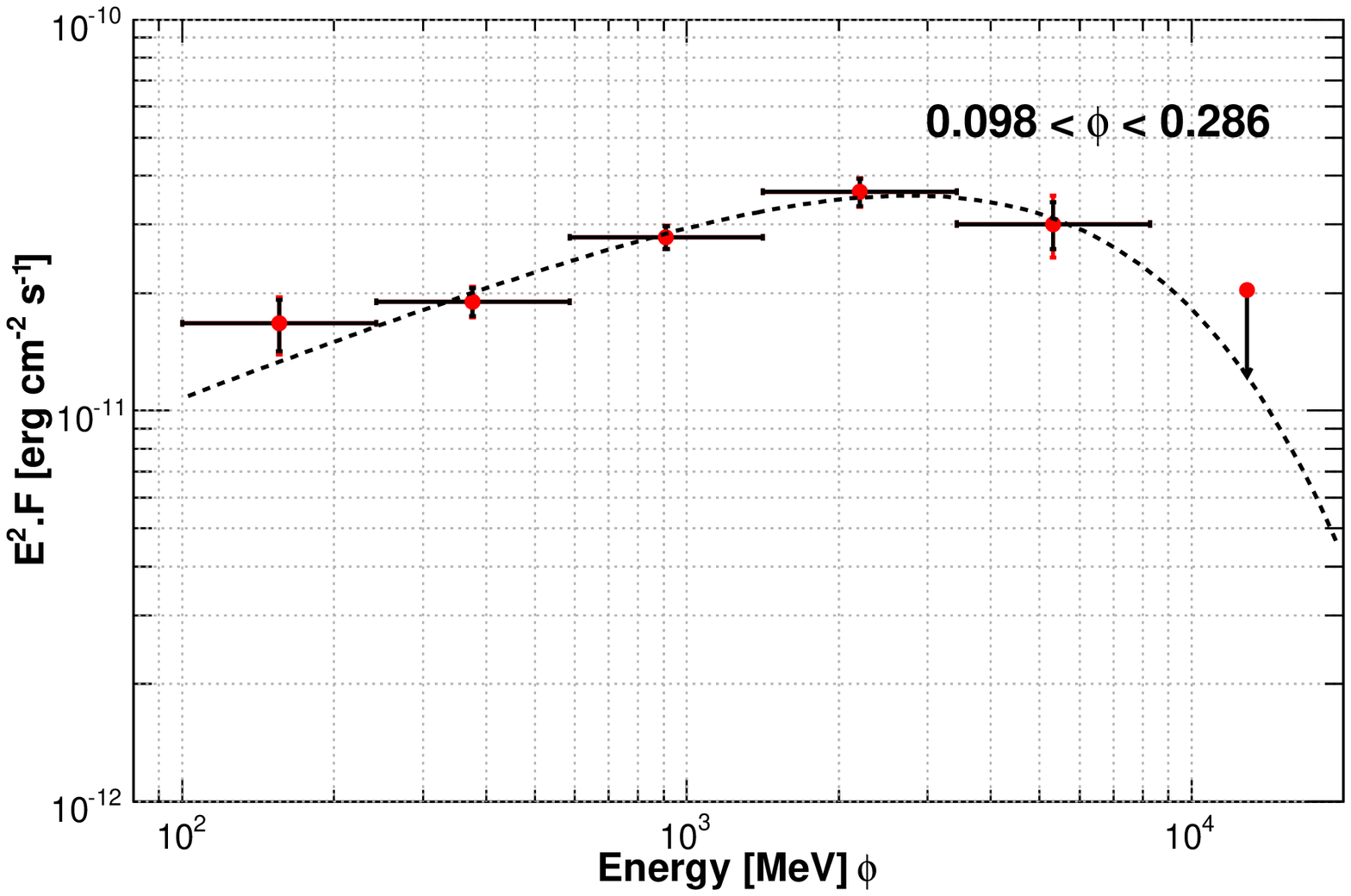}
\plotone{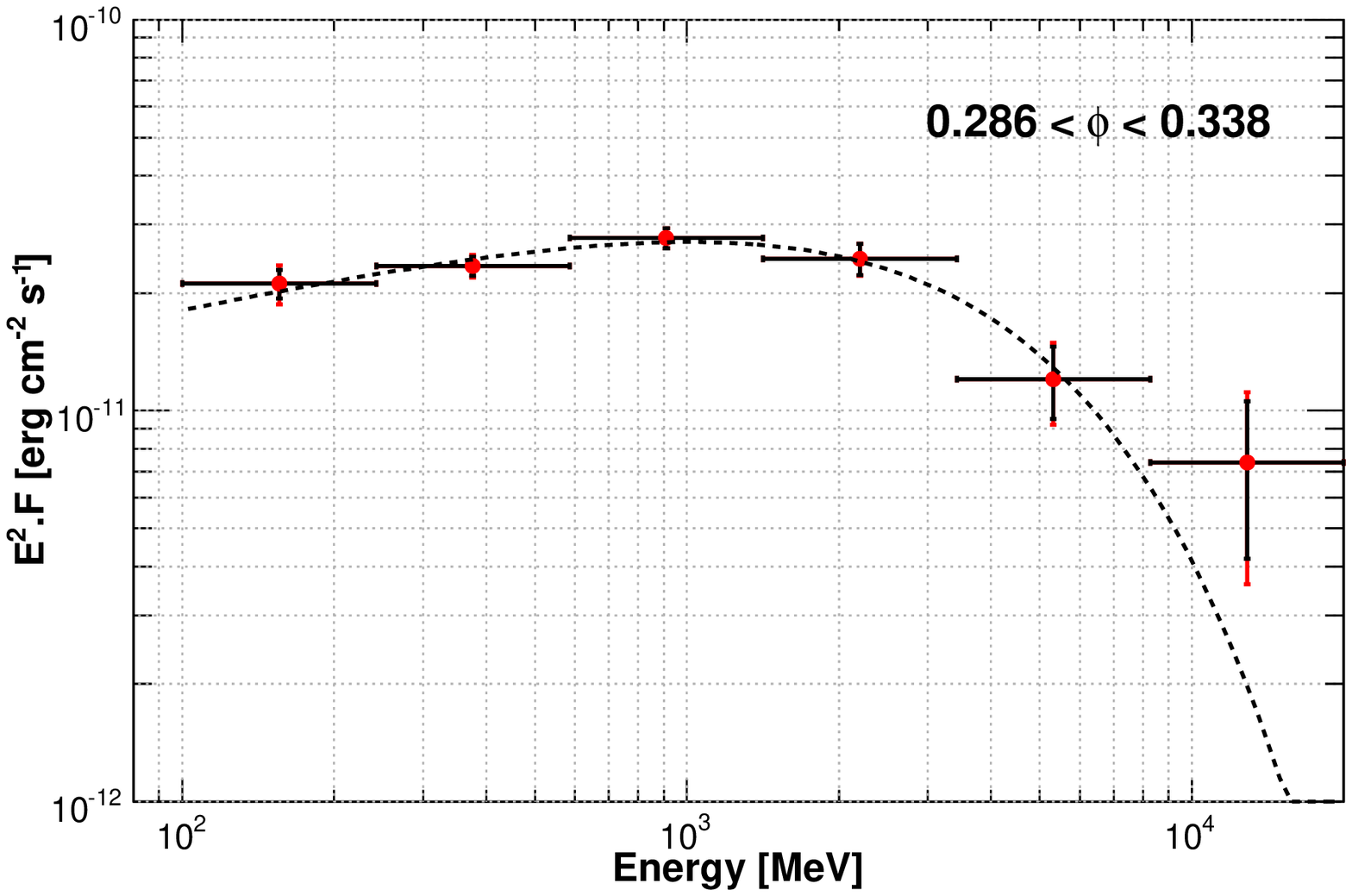}
\end{minipage} \hfill
\begin{minipage}[c]{.98\linewidth}
\epsscale{.58}
\plotone{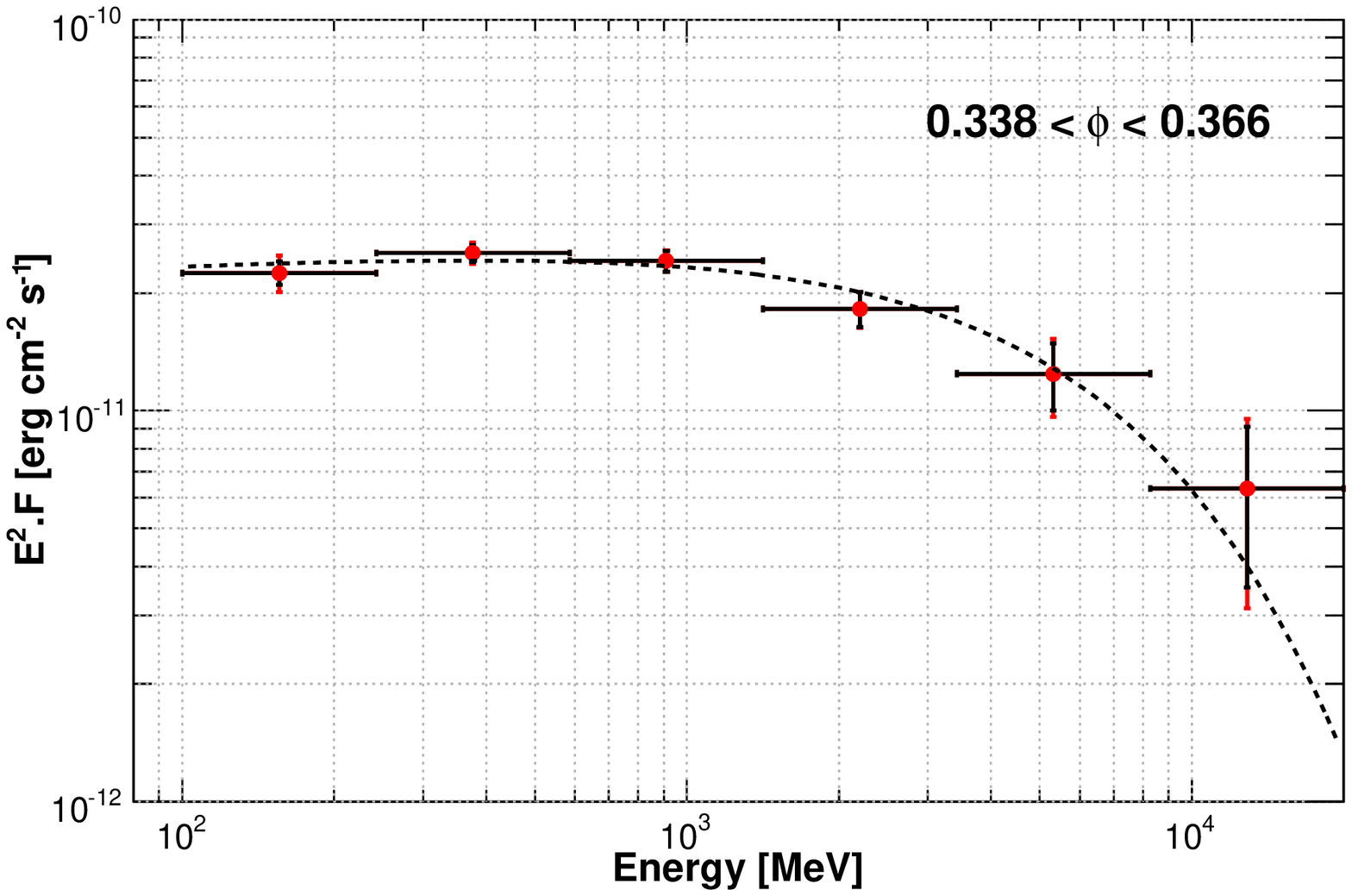}
\plotone{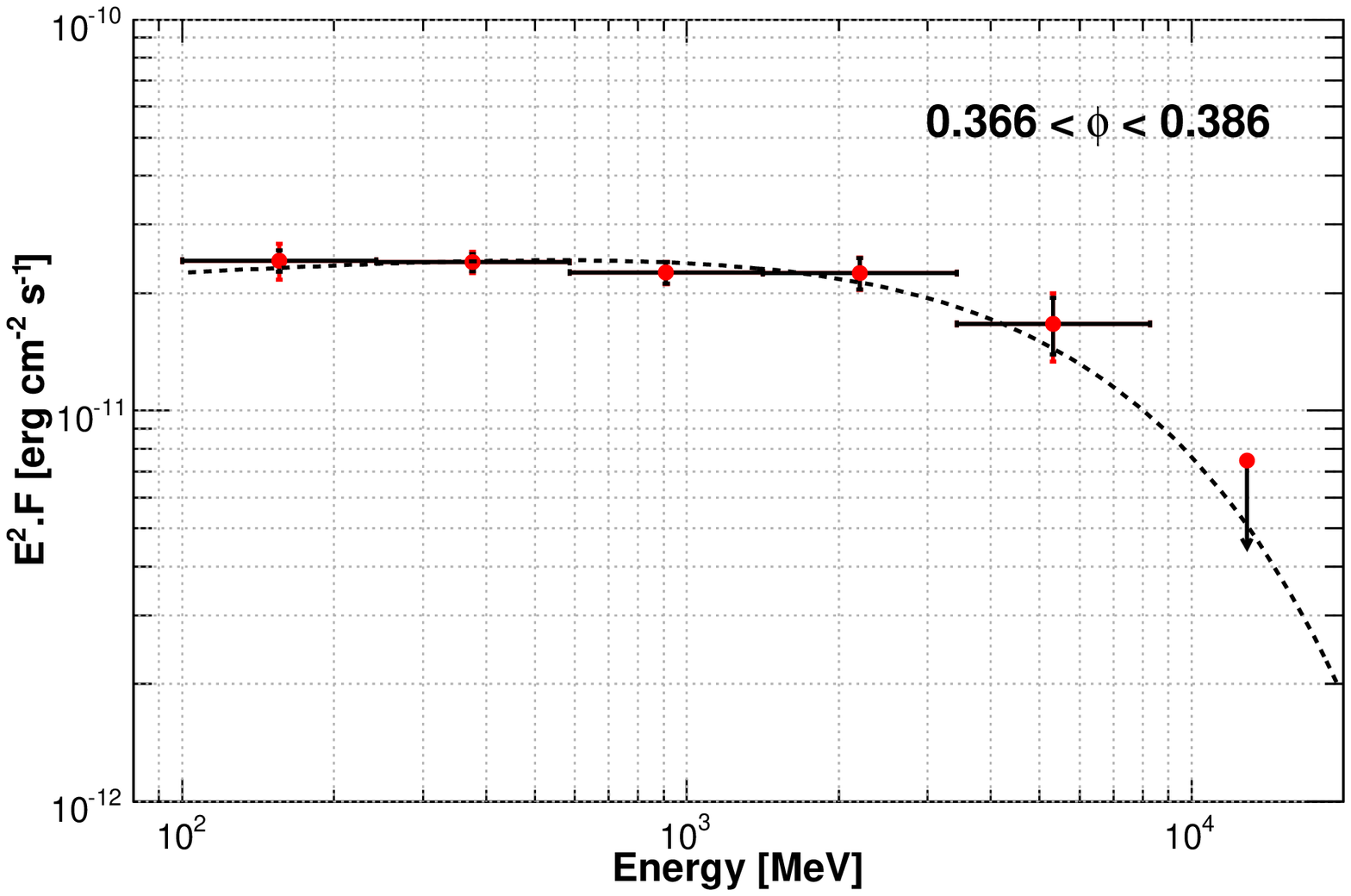}
\end{minipage} \hfill
\begin{minipage}[c]{.98\linewidth}
\epsscale{.58}
\plotone{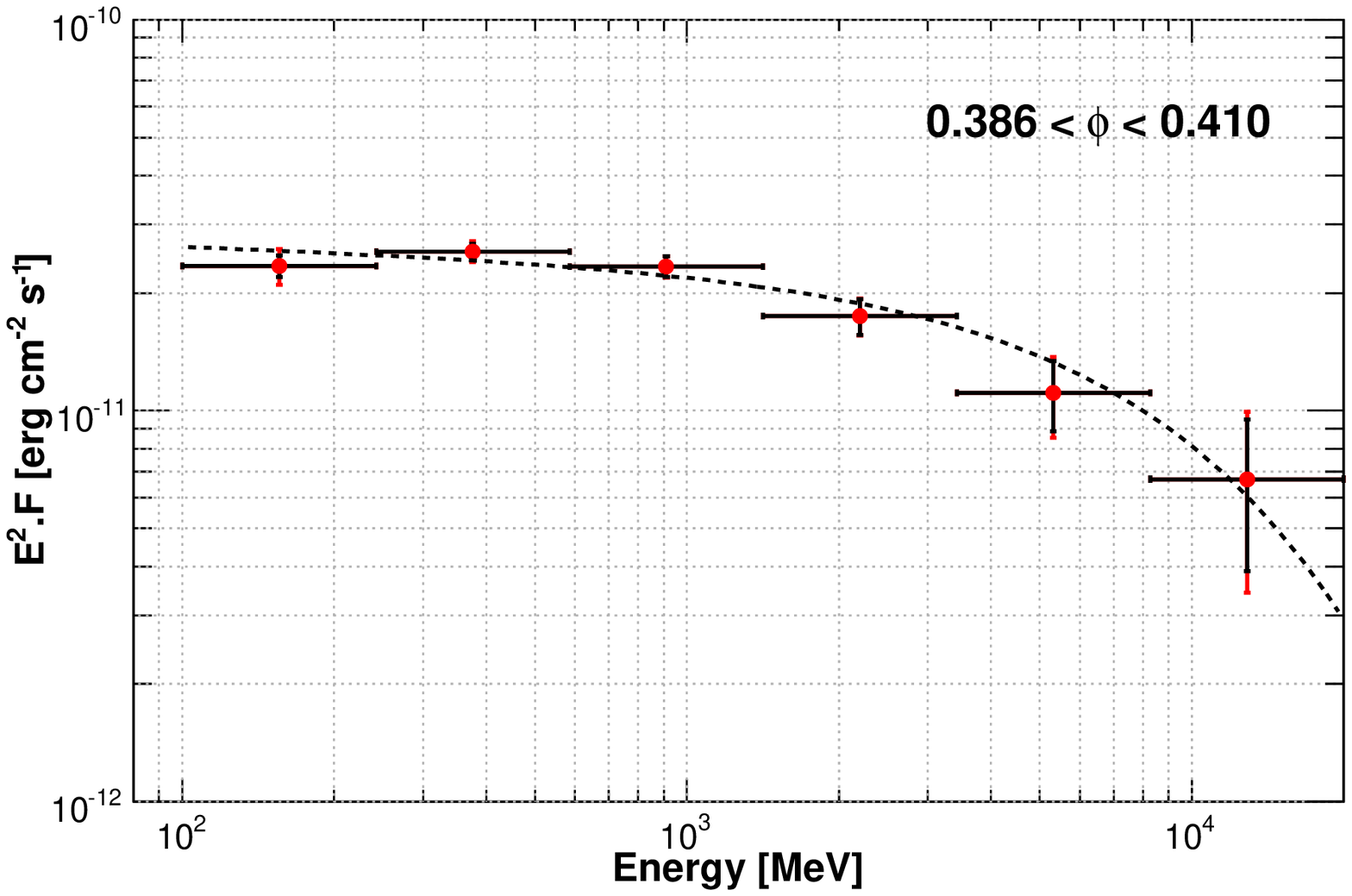}
\plotone{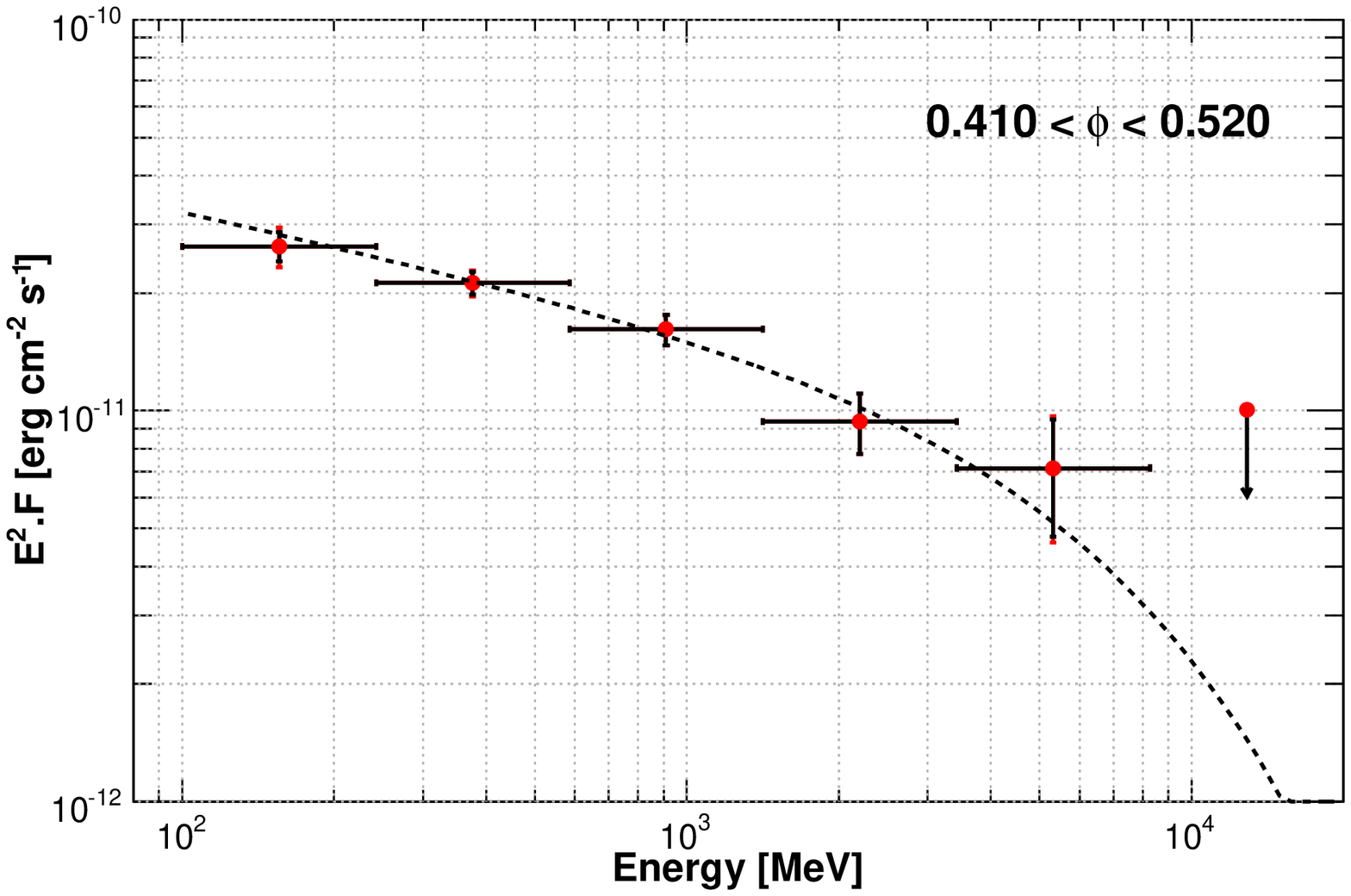}
\end{minipage} \hfill
\caption{\label{resultats_phase_resolved}Phase-resolved spectral energy distributions of the Crab Pulsar. The labels indicate the phase intervals. Spectral results are presented in Table~\ref{table_phase_resolved}. The black dotted curve is the best fit power-law with an exponential cut-off. The LAT spectral points (cf. Figure~\ref{SED_neb_8mois} for the description of the conventions) are obtained using the model-independent maximum likelihood method described in Section~\ref{nebula}. 90~\% C.L. upper limits are computed when the statistical significance is lower than 3~$\sigma$.}
\end{figure*}

\begin{figure*}[h!]
\epsscale{1.10}
\plotone{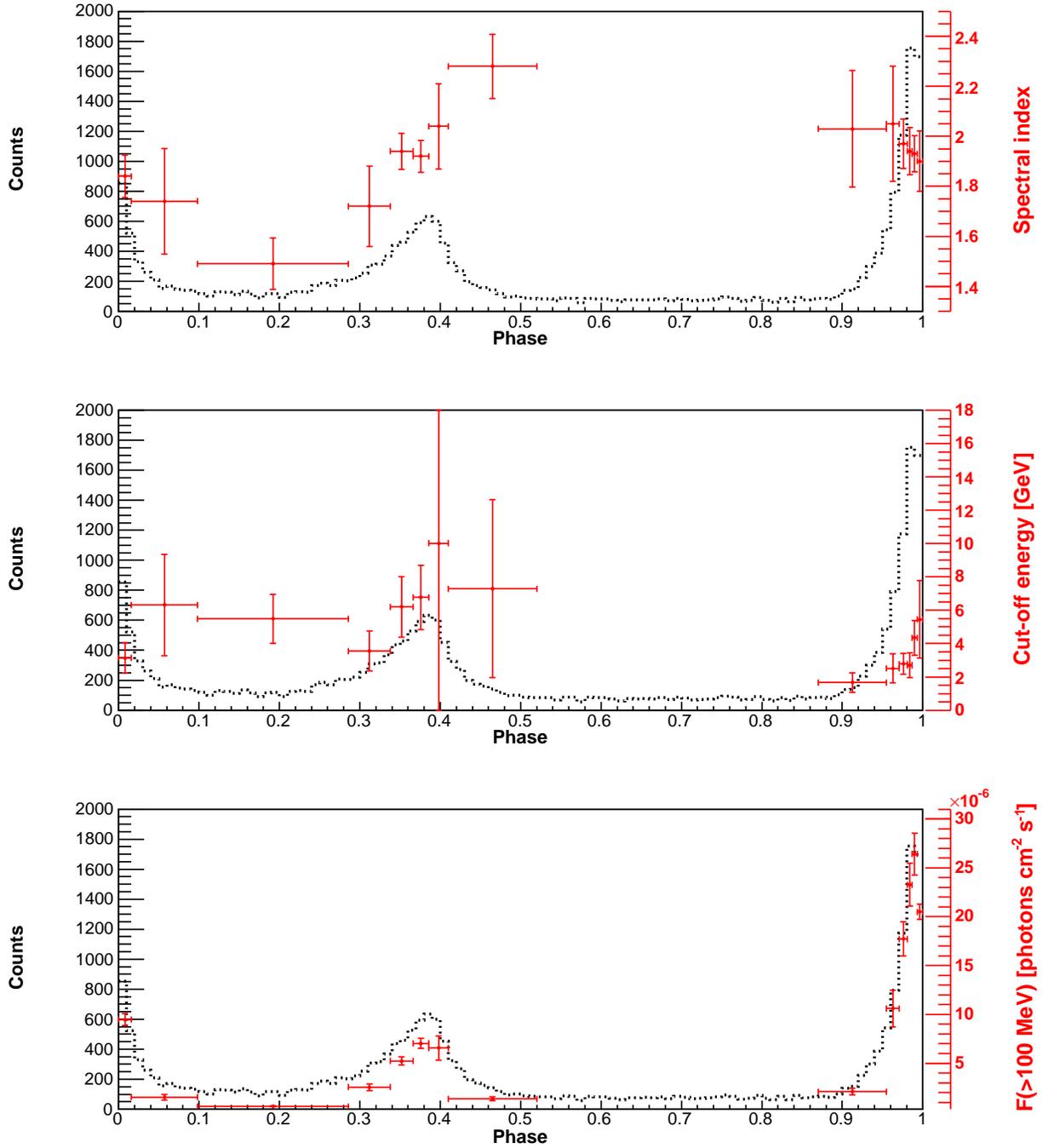}
\caption{\label{phase_resolved_parameters}Variation of the spectral indices, cut-off energies and photon flux above 100 MeV (divided by the phase interval width) as the function of the pulse phase. A power-law with an exponential cut-off shape has been assumed for each phase interval (defined in Table~\ref{table_phase_resolved}). Vertical bars show the combined statistical and systematic errors. The horizontal bars delimit the phase interval containing $\sim 1000$ pulsed photons in the energy-dependent region defined in Section~\ref{phasos}. The dotted histogram represents the \emph{Fermi}-LAT lightcurve above 100 MeV, binned to 0.01 in phase.}
\end{figure*}

\begin{figure*}[h!]
\begin{center}
\epsscale{1.2}
\plotone{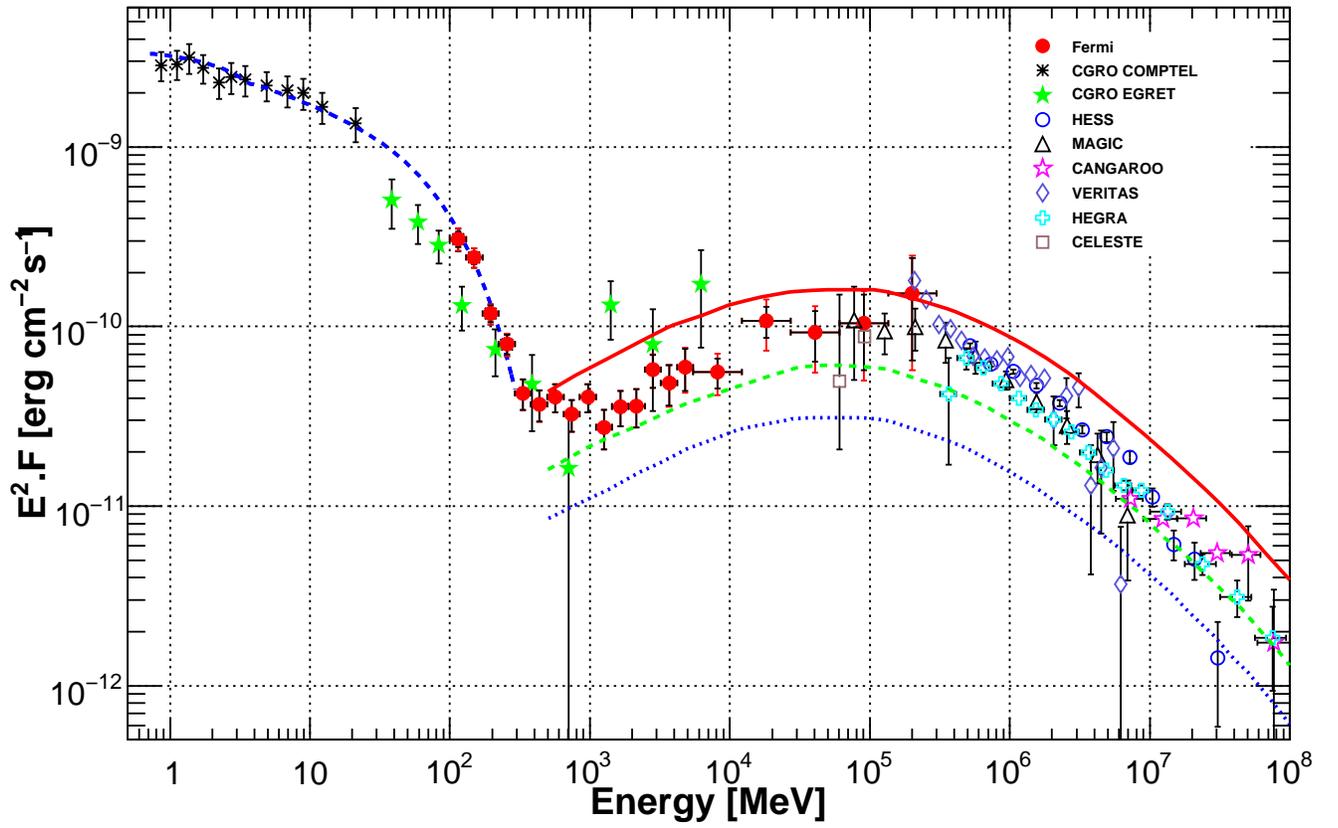}
\caption{\label{SED_nebula}The spectral energy distribution of the Crab Nebula from soft to very high energy $\gamma$-rays. The fit of the synchrotron component, using COMPTEL and LAT data (blue dashed line), is overlaid. The predicted inverse Compton spectra from \cite{Atoyan and Aharonian 1996} are overlaid for three different values of the mean magnetic field: 100~$\mu$G (solid red line), 200~$\mu$G (dashed green line) and the canonical equipartition field of the Crab Nebula 300~$\mu$G (dotted blue line). References: CGRO COMPTEL and EGRET: \cite{Kuiper et al. 2001}; MAGIC: \cite{Albert et al. 2008}; HESS: \cite{Aharonian et al. 2006}; CANGAROO: \cite{Tanimori et al. 1997}; VERITAS: \cite{Celik 2007}; \textbf{HEGRA: \cite{Aharonian et al. 2004}; CELESTE: \cite{Smith et al. 2006}}}
\end{center}
\end{figure*}

\end{document}